\documentclass[10pt, aps, pra, preprintnumbers,nofootinbib]{revtex4-2}
\usepackage{graphicx}
\usepackage{dcolumn}
\usepackage{amsmath}
\usepackage{float}
\usepackage{makecell}
\usepackage{bm}
\usepackage{comment}
\usepackage{subcaption}
\usepackage{slashed}
\usepackage{gensymb}
\usepackage{tikz}
\usepackage[normalem]{ulem}
\usepackage[compat=1.1.0]{tikz-feynman}
\usepackage{ragged2e}
\usepackage[utf8]{inputenc}
\usepackage{tabularx}
\usepackage{array}
\usepackage{graphics}
\graphicspath{{}}
\usepackage{psfrag}
\usepackage{epsfig}
\usepackage{amssymb}
\usepackage{setspace}
\usepackage{rotating}
\usepackage{colortbl}
\usepackage{slashed}
\usepackage{enumitem}
\makeatletter
\usepackage{textcomp}
\usepackage[colorlinks=true,allcolors=purple]{hyperref}
\usepackage[capitalize]{cleveref}

\definecolor{nicegreen}{rgb}{0., 0.75, 0.46}

\begin{document}

\begin{flushright}
MI-HET-842
\end{flushright}

\title{Dirt/Detector/Dump:\\ Complementary BSM production at Short-Baseline Neutrino Facilities}

\author{Bhaskar Dutta}
\email{dutta@tamu.edu}
\affiliation{
Mitchell Institute for Fundamental Physics and Astronomy,
Department of Physics and Astronomy, Texas A\&M University, College Station, TX 77843, USA
}
\author{Debopam Goswami}
\email{debopam22@tamu.edu}
\affiliation{
Mitchell Institute for Fundamental Physics and Astronomy,
Department of Physics and Astronomy, Texas A\&M University, College Station, TX 77843, USA
}
\author{Aparajitha Karthikeyan}
\email{aparajitha\_96@tamu.edu}
\affiliation{
Mitchell Institute for Fundamental Physics and Astronomy,
Department of Physics and Astronomy, Texas A\&M University, College Station, TX 77843, USA
}
\author{Kevin J. Kelly}
\email{kjkelly@tamu.edu}
\affiliation{
Mitchell Institute for Fundamental Physics and Astronomy,
Department of Physics and Astronomy, Texas A\&M University, College Station, TX 77843, USA
}

\date{\today}

\begin{abstract}
Short-baseline neutrino (SBN) facilities are optimal for new-physics searches, including the possible production of new particles in and along the neutrino beamline. One such class of models considers states that are created by neutrino upscattering that then decay in the neutrino detector -- in the past, such upscattering has often been considered to occur in the detector itself (with a prompt decay) or in the dirt upstream of the detector. In this work, we highlight the importance of the beam dumps, situated even further upstream, for such searches. The Fermilab Booster Neutrino Beam, with its iron dump, provides one such possibility. We focus on sub-GeV heavy neutral leptons (HNLs) with a transition magnetic moment, which allows this upscattering to take advantage of the high-$Z$ iron. We observe that, in addition to increased sensitivity to this model at SBND, MicroBooNE, and ICARUS, there exist distinct features in the signal events' kinematical properties when coming from production in the dump, dirt, and detector which can allow for enhanced signal-to-background separation. We highlight the complementarity of this approach to study parameter space relevant for the MiniBooNE low-energy excess, as well as in models in which the HNLs couple to a light scalar particle.\footnote{The work and conclusions presented in this manuscript are not intended to be interpreted as official results from the SBND collaboration.}
\end{abstract}

\maketitle

\section{\label{sec:level1}Introduction}

Despite the considerable success of the Standard Model (SM) of particle physics in the incorporation of neutrino interactions, it remains incomplete in accounting for neutrino mass: within the SM, neutrinos are massless. 
However, this is inconsistent with many past experiments~\cite{Davis:1968cp, Abazov:1991rx, GALLEX:1992gcp, Kamiokande:1996qmi, Super-Kamiokande:1998kpq, SNO:2002tuh} which have demonstrated that neutrinos have mass, providing unequivocal evidence for physics beyond the SM.

In the pursuit of understanding why neutrinos have such small masses, various theoretical explanations have been proposed, with the see-saw mechanism~\cite{Mohapatra:1979ia, Schechter:1980gr} being particularly compelling. In the low-scale seesaw mechanism, additional neutrino mass eigenstates are postulated. The relatively small mass of neutrinos compared to other fermions and the distinct structure of leptonic mixing, as opposed to that of quarks, suggest that neutrino masses might originate from a mechanism different from the Higgs mechanism. In many scenarios, right-handed neutrinos are crucial for explaining neutrino masses, and when their masses are significantly larger than $\mathcal{O}(\text{eV})$, they are referred to as Heavy Neutral Leptons (HNLs).

While there are many models in which a HNL can appear (see, e.g., Ref.~\cite{Abdullahi:2022jlv} and references therein), we focus on those that interact with neutrino flavor eigenstates via a transition magnetic moment (dipole portal) operator~\cite{Shrock:1982sc, Magill:2018jla, Ovchynnikov:2023wgg, Barducci:2023hzo, Shoemaker:2018vii, Dasgupta:2021ies, Masip:2012ke, Ismail:2021dyp, Shoemaker:2020kji, Brdar:2020quo, Coloma:2017ppo, Batell:2022xau, Huang:2022pce}. Such a dipole portal HNL (dpHNL) is of great interest as it serves as a possible contribution to the excess of electron-like events that were observed in the MiniBooNE (MB) experiment~\cite{MiniBooNE:2018esg, MiniBooNE:2020pnu}, alongside contributions from a light sterile neutrino impacting oscillations. This explanation has notable advantages, including a better fit to the low-energy and forward-angle 
excess, and also alleviating tensions in global fits to the 3+1 model~\cite{Kamp:2022bpt}, where one sterile neutrino is introduced to the three active neutrinos already present.

In a typical beam dump experiment (for example, the 8~GeV Fermilab Booster Neutrino Beam (BNB)~\cite{MicroBooNE:2015bmn, Machado:2019oxb}), such dpHNLs are dominantly produced via upscattering when SM neutrinos propagate through the detector, the surrounding material (dirt), and/or the 
dump. While the contributions from the first two sources have been considered in previous studies~\cite{Kamp:2022bpt, Shoemaker:2018vii}, the 
contribution from the dump has not been examined. 

In this study, we find that including dpHNLs created in the dump can enhance the sensitivity, particularly at closer detectors such as the Short Baseline Near Detector (SBND)~\cite{MicroBooNE:2015bmn}. We also examine the sensitivities at MicroBooNE~\cite{MicroBooNE:2016pwy}, MiniBooNE~\cite{MiniBooNE:2008hfu, MiniBooNE:2008paa}, and ICARUS~\cite{ICARUS:2004wqc}, which are located further away along the BNB beamline. Additionally, we identify distinguishing features in the energy, angular, and timing spectra~\cite{Shrock:1978ft} based on the dpHNL's point of production. We propose that the study of these features can be utilized for a detailed background analysis to obtain a comprehensive sensitivity. We further consider the dpHNL scenario proposed in Refs.~\cite{Vergani:2021tgc, Kamp:2022bpt} as a solution to the MiniBooNE anomaly and examine its possible detection at SBND. We broaden our study by including HNLs which interact with SM neutrinos via a sub-GeV massive scalar~\cite{Dutta:2020scq, Falk:1999mq}. Such a model also has the potential to explain the MiniBooNE anomaly~\cite{Dutta:2020scq}. Therefore, we also look at signals from neutrino upscattering that are facilitated by this massive scalar mediator. Following this, we will refer to the transition magnetic moment scenario as the dpHNL model and the one with the massive scalar as the Light Scalar Mediator (LSM) model. 
 
This paper is organized as follows: In Section~\ref{SectionII}, we briefly discuss the two models we consider in our study. We also present the critical kinematic equations, cross sections for the (dpHNL) production, and detection mechanisms. The experimental details are outlined in Section~\ref{SectionIII}. Our results are presented in Sections~\ref{sec:Distributions}, \ref{sec:SignalVsBackground}, and \ref{sec:ResultsDiscussion}, followed by our conclusions in Section~\ref{SectionV}. Appendix~\ref{AppendixA} provides the fluxes used in our analysis, and Appendix~\ref{AppendixB} elaborates on the exact kinematics of our calculations followed by Appendix~\ref{AppendixC} which describes the cross-sections for different processes and the decay widths used. Appendix~\ref{app:MiniBooNEExcessKinematics} presents the energy and angular spectra at SBND for the dpHNL model parameters used to explain the MiniBooNE excess. Additional sensitivity plots for MicroBooNE and ICARUS are provided in Appendix~\ref{app:MoreSensitivity}. To facilitate a clearer comparison of the spectra as a function of distance from the target, all corresponding distributions for ICARUS are included in Appendix~\ref{app:MoreComparisons}.

\section{Models}\label{SectionII}
Generally, we are interested in signatures of particles that are capable of being produced from neutrino upscattering off SM targets and can travel macroscopic distances before decaying back to SM signatures. Heavy Neutral Leptons (HNLs) are a frequently studied class of models where the ${\sim}$MeV-GeV scale fermions, through their small couplings to SM neutrinos and thanks to weak-interaction suppression, have relatively long lifetimes. Depending on the HNL mass and the structure of its interactions with the SM, it can decay into a variety of visible SM final states -- in this work, we will be particularly interested in decays into photons or into electron/positron pairs.

We focus on two scenarios involving HNLs, each with a key ingredient that allows for efficient production in a neutrino beam. They are also associated with common solutions to the low-energy excess observed by the MiniBooNE experiment~\cite{MiniBooNE:2018esg}. In~\cref{subsec:dpHNL}, we explore the scenario in which the HNLs couple to the SM via a transition magnetic moment, the so-called dipole-portal HNLs (dpHNL). The majority of our analyses to follow will focus on the dpHNL scenario. Alternatively,~\cref{subsec:LSMModel} presents a scenario in which the HNLs couple to the SM (and decay to electron/positron pairs) via a new light scalar particle.

\subsection{Dipole-coupled Heavy Neutral Leptons}\label{subsec:dpHNL}
The effective Lagrangian of a dipole-coupled Heavy Neutral Lepton (dpHNL)~\cite{Magill:2018jla} can be expressed as
\begin{equation}\label{1}
\mathcal{L}_{\text{dpHNL}} \supset \bar{N}(i\slashed{\partial}-m_N)N + d_\mu(\bar{\nu}_L \sigma_{\lambda \rho}F^{\lambda \rho}N) + \mathrm{h.c.,}
\end{equation}
where $m_N$ is the mass of the dpHNL $N$, $F_{\mu \nu}$ is the SM electromagnetic field strength tensor. We label the dipole coupling as $d_\mu$ to specify that we focus on coupling with SM neutrinos in the $\mu$ flavor state. In this work, we assume the dpHNL to be purely Dirac. The coupling $d_\mu$ carries mass-dimension $-1$ and the lack of gauge invariance of~\cref{1} indicates some necessary UV completion. Nevertheless, it serves as a valid, effective Lagrangian at the scales of interest in accelerator neutrino environments (${\sim}$GeV and below) such as the SBN program~\cite{MicroBooNE:2015bmn}.

\subsubsection{Production of dpHNLs}
\begin{figure}[!htbp]
\centering
    \begin{subfigure}{0.35\textwidth}
        \begin{tikzpicture} 
            \begin{feynman} 
            \vertex(a0);\vertex[right=of a0](a1); \vertex[right=of a1](a2); \vertex[below=of a1](a3); \vertex[left=of a3](a4); \vertex[right=of a3](a5);
                
                \diagram*{ (a0)--[fermion, edge label'=\(\nu_{\mu}\)](a1),(a1)--[fermion, edge label'=\(N\)](a2),(a1)--[photon, edge label=\(\gamma\)](a3),(a4)--[fermion, edge label'=\(T\)](a3), (a3)--[fermion, edge label'=\(T\)](a5)}; 
            \end{feynman}
        \end{tikzpicture}
    \caption{}
    \label{fig:upscatter}
    \end{subfigure}
    \begin{subfigure}{0.35\textwidth}
        \begin{tikzpicture} 
            \begin{feynman} 
                \vertex(a0);\vertex[right=of a0](a1); \vertex[above right=of a1](a2); \vertex[below right=of a1](a3);
                
                \diagram*{ (a0)--[fermion, edge label'=\(N\)](a1),(a1)--[fermion, edge label'=\(\nu_\mu\)](a2),(a1)--[photon, edge label=\(\gamma\)](a3)}; 
            \end{feynman}
        \end{tikzpicture}
    \caption{}
    \label{fig:decay}
    \end{subfigure}
    \captionsetup{justification=Justified,singlelinecheck=false}
    \caption{Feynman Diagrams of (left): Primakoff scattering of $\nu_\mu$ with target nuclei $T$, resulting in the production of a dpHNL ($N$) mediated by a SM photon, and (right): Decay of the dpHNL into $\nu_\mu$ and $\gamma$.}
\end{figure}
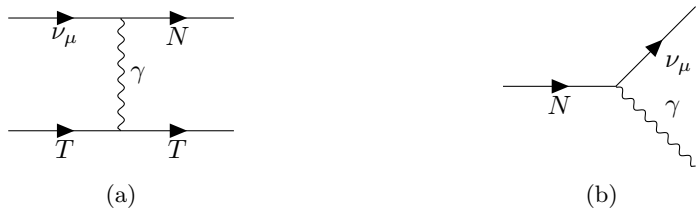

In accelerator-neutrino beam environments, the most efficient\footnote{Another possibility is the production of $N$ in three-body charged-meson decays, e.g., $\pi^+ \to \mu^+ N \gamma$~\cite{Barducci:2023hzo}. This process is subdominant for all parameter spaces of interest in this work.} production of dpHNLs comes from neutrino upscattering when the SM neutrinos interact with a target nucleus (see~\cref{fig:upscatter}, commonly referred to as Primakoff scattering~\cite{Primakoff:1951iae}). This scattering process, as it is mediated by the SM photon, prefers low-momentum-transfer scattering and is therefore enhanced by the target nucleus's atomic number $Z$~\cite{Tsai:1986tx}. In the remainder of this section, we will briefly explain the differential cross-section of the Primakoff process, the kinematics of the dpHNLs produced, and obtain an expression for the dpHNL flux.

We choose to express the differential upscattering cross section, with respect to the outgoing target's recoil energy $E_R$, which can be expressed as
\begin{widetext}
\begin{equation}\label{eq:UpscatteringCrossSection}
\frac{d\sigma_{(\nu_\mu T \rightarrow N T)}}{dE_R} = 4d_\mu^2 \alpha_{\rm EM} Z^2 \left \lvert F(E_R)\right\rvert^2 \left(\frac{1}{E_R} - \frac{m_N^2}{2E_\nu E_R m_T}\Bigg(1-\frac{E_R}{2E_\nu} + \frac{m_T}{2E_\nu}\Bigg)-\frac{1}{E_\nu} + \frac{m_N^4(E_R - m_T)}{8E_\nu^2 E_R^2 m_T^2}\right).
\end{equation}
\end{widetext}
Here, $m_T$ and $Z$ are the target nucleus's mass and atomic number, respectively, and $E_\nu$ is the incoming neutrino energy. The conservation of energy dictates that the outgoing dpHNL energy is $E_N = E_\nu - E_R$. Kinematically, the minimum neutrino energy required to produce dpHNL is $E_\nu^{\rm min.} = m_N + m_N^2/(2m_T)$.
Our calculation agrees with Refs.~\cite{Shoemaker:2018vii, Brdar:2020quo}. We parameterize the nuclear form factor $F(E_R)$ using~\cite{Helm:1956zz, Engel:1991wq}
\begin{equation}\label{10a}
F(E_R) = \frac{3}{(\kappa r)^3}e^{-\kappa^2 s^2 /2}(\sin(\kappa r) - \kappa r \cos(\kappa r)),
\end{equation}
where $s=0.9~\text{fm}$, $r = 3.9 \times (A/40)^{1/3}~\text{fm}$ and, $\kappa=\sqrt{E_R^2+2m_N E_R}$. We neglect the nuclear magnetic form factor, as it does not benefit from the $Z^2$ enhancement present here~\cite{Brdar:2020quo}. 

Full kinematical details of this process are provided in~\cref{AppendixB} -- here, we provide some key features. First, the range of allowed recoil energies for the nucleus is specified by
\begin{widetext}
\begin{equation}\label{eq:ERMinMax}
E_{R}^{\text{min/max}} = \frac{2m_T E_\nu^2 - m_N^2(E_\nu +m_T) \mp E_\nu\sqrt{4m_T^2 E_\nu^2 - 4m_T m_N^2(E_\nu + m_T) + m_N^4}}{2m_T(2E_\nu +m_T)},
\end{equation}
\end{widetext}
which reduces to the standard coherent elastic neutrino scattering result as $m_N \rightarrow 0$~\cite{Brdar:2018qqj}. Additionally, the angle of the outgoing dpHNL relative to the incoming neutrino, $\theta_N$, is restricted given the incident/outgoing particle energies,
\begin{equation}\label{13a}
\cos \theta_N=\frac{1}{E_\nu p_N}\Bigg(E_\nu E_N - \frac{m_N^2+2m_T E_R}{2}\Bigg).
\end{equation}
From these expressions, we can understand that the differential cross section favors $E_R \to 0$ (originating from $t \to 0$ in the $t$-exchange diagram apparent in~\cref{fig:upscatter}). In this case, $E_N \to E_\nu$ and $\cos\theta_N \to 1$; the dpHNL will be predominantly forward-going and carry the bulk of the incoming neutrino energy.

We can calculate the flux of dpHNL, given an incoming flux of $\nu_\mu$, using
\begin{equation}\label{eq:dpHNLFlux}
\frac{d\Phi_N}{dE_N} = N_T \int_{E_\nu^{\rm min.}}^{\infty} \frac{d\Phi_\nu}{dE_\nu} \frac{d\sigma}{dE_N} dE_\nu,
\end{equation}
where $N_T$ is the number of nuclear targets available for upscattering and we have transformed the differential cross section to be with respect to $E_N$ instead of $E_R$ (with a trivial Jacobian factor).

\subsubsection{Decay \& Detection of dpHNLs}
Through the same interaction with which they are produced, dpHNLs are unstable and can decay into a SM neutrino and a photon (see~\cref{fig:decay}) with 100\% branching ratio. If such a decay occurs within a neutrino detector and the photon energy is sufficiently high, it will be identifiable as an event. We will discuss signal identification and separation from background events in detail in the following. The rest-frame decay width of the dpHNL $\Gamma_N$, as well as the lab-frame decay-length $\lambda_N$, are
\begin{equation}\label{16}
\Gamma_N = \frac{d_\mu^2 m_N^3}{4\pi}, \quad \lambda_N = \gamma \beta c\tau_N.
\end{equation}

For the dpHNL to reach the detector and decay within, it must be sufficiently boosted. Assuming that the dpHNL is produced at a distance $d > 0$ upstream\footnote{For an upscattering-produced dpHNL that is generated inside the detector, one may take the $d\to0$ limit of~\cref{eq:Pdecay}, and substitute $L$ with the distance between the dpHNL's production point and the edge of the detector, along its path of travel.} of the detector, along a path that intersects the detector with a length $L$, the probability of this decay occurring is
\begin{equation}\label{eq:Pdecay}
P_{\rm decay} = e^{-d/\lambda_N}\left(1 - e^{-L/\lambda_N}\right).
\end{equation}
Combining this with the dpHNL flux given in~\cref{eq:dpHNLFlux}, we may estimate the expected signal event rate $R$ as
\begin{equation}\label{eq:dpHNLEventRate}
R = \int_{E_{N}^{\rm min.}}^{E_N^{\rm max.}} \frac{d\Phi_N}{dE_N} P_{\rm decay} dE_N,
\end{equation}
where any energy-dependent efficiency (e.g., in terms of the outgoing photon energy) may be incorporated in the integrand. In this work, we consider 100\% efficiency.

\subsection{Light-Scalar Mediated HNLs}\label{subsec:LSMModel}
A similar HNL scenario exhibiting the same features we are interested in here is one where the HNL couples to the SM via a light scalar $h_1$ instead of the dipole-portal coupling $d_\mu$. The scalar can also couple to SM fermions via Yukawa couplings, given the Lagrangian
\begin{equation}\label{1.28}
\mathcal{L} \supset (y_{nh_1})_{22} h_1 \bar{N} \nu_\mu + \sum_f (y_{fh_1})_{11}\bar{f}fh_{1} + h.c.
\end{equation} 
The scalar has a mass $m_{h_1}$, leading to differences in the production kinematics of the HNLs in the neutrino beam. If $y_{eh_1}$, the Yukawa coupling between $h_1$ and SM electrons is nonzero, then this scenario offers a potential explanation to the MiniBooNE low-energy excess~\cite{Dutta:2020scq}, with $h_1 \to e^+ e^-$ decays occurring within the detector. We will focus on this scenario, in which neutrino upscattering (mediated by $h_1$) produces $N$ in the beam, and then $N$ decays to $e^+ e^-$ pairs via sequential $N \to \nu h_1$, $h_1 \to e^+ e^-$ decays.

We provide detailed calculations of the upscattering cross section akin to~\cref{eq:UpscatteringCrossSection} in~\cref{AppendixC}. 
The kinematics are similar to the dpHNL scenario; however, the events no
longer favor $E_R \to 0$ as strongly, given the nonzero $h_1$ mediator
mass.

\section{SBN Experimental Details}\label{SectionIII}

\begin{figure*}[!htbp]
    \centering
        \includegraphics[scale=1.0]{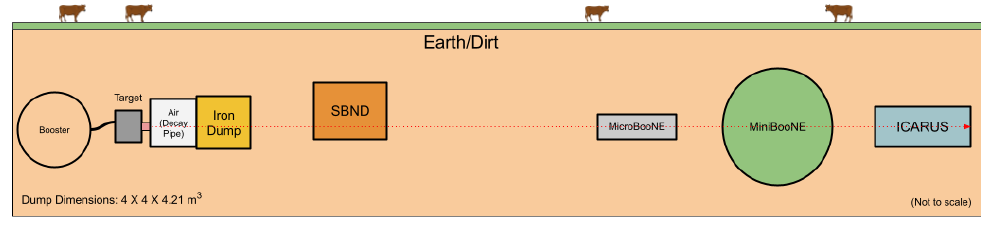}
    \caption{Schematic Diagram of the SBN Experimental Program (not to scale).} 
    \label{fig:1.1}
\end{figure*}

A schematic diagram of the SBN experimental program~\cite{Machado:2019oxb, MicroBooNE:2015bmn} is presented in Fig.~\ref{fig:1.1}. The specific experimental parameters used in this analysis are detailed in~\cref{tab:ExperimentalDetails}. Let us briefly summarize the key features of the SBN facilities that are of interest for this study. The Booster Neutrino Beam is initiated with $8$~GeV protons striking a beryllium target; the magnetic focusing horns lead to a nearly pure $\nu_\mu$ (or $\bar\nu_\mu$) beam pointing in the forward direction, with the neutrinos produced throughout the decay pipe. At the end of this decay volume, a $4\times 4\times 4.21$~m$^3$ iron dump stops any charged SM particles still traveling in the forward direction, ensuring that only neutrinos reach the downstream detectors. We discuss the fluxes in detail in~\cref{AppendixA}, following studies in Ref.~\cite{MicroBooNE:2015bmn}.

SBN features three liquid-argon time-projection chamber (LArTPC) detectors: SBND~\cite{MicroBooNE:2015bmn}, MicroBooNE~\cite{MicroBooNE:2016pwy}, and ICARUS~\cite{ICARUS:2004wqc}, located at 110~m, 470~m, and 600~m from the BNB target, respectively. SBND and ICARUS are operational, while MicroBooNE has completed taking data. We will also study the MiniBooNE detector, which was situated 537~m from the BNB target and consisted of an 818~t mineral-oil (CH$_2$) Cerenkov detector.

\begin{table*}[!htbp]
    \centering
    \renewcommand{\arraystretch}{1.2}
    \begin{tabular}{|c|c|c|c|c|c|c|c|}
        \hline
        Experiment & \makecell{Mass\\(tons)} & \makecell{Dimensions\\(m $\times$ m $\times$ m)} & \makecell{Distance\\(m)} & \makecell{Total\\POT} & \makecell{Background\\Counts} & \makecell{Angle off-axis\\(degrees)}& \makecell{Threshold\\Energy\\(MeV)}\\
        \hline
        SBND~\cite{MicroBooNE:2015bmn} & 112.0 & $4 \times 4 \times 5$ & 110 & $10^{21}$ & 9000 & 0.3 & 30\\
        (LAr, Ongoing) & & & & & & &  \\
        \hline
        MicroBooNE~\cite{MicroBooNE:2016pwy} & 86.8 & \makecell{$2.6 \times 2.3 \times$\\$10.4$} & 470 & \makecell{$6.6 \times 10^{20}$} & 167~\cite{Mogan:2021iau} & 0 & 30\\
        (LAr, Completed) & & & & & & &  \\
        \hline
        MiniBooNE~\cite{MiniBooNE:2008hfu} & 818.0 & \makecell{Radius = 5} & 537 & \makecell{$1.875 \times 10^{21}$} & 2215~\cite{MiniBooNE:2020pnu} & 0 & 100\\
        (Mineral Oil $\text{CH}_2$, Completed) & & & & & & &  \\
        \hline
        ICARUS T600~\cite{ICARUS:2004wqc} & 476.0 & \makecell{$3.6 \times 3.9 \times$\\$19.6$} & 600 & $10^{21}$ & 840 & 0 & 30\\
        (LAr, Ongoing) & & & & & & &  \\
        \hline
    \end{tabular}
    \caption{SBN experimental specifications utilized in this analysis~\cite{ICARUS:2023gpo, 
Machado:2019oxb, PRISM}.} 
    \label{tab:ExperimentalDetails}
\end{table*}

For all detectors, we will study three different production scenarios for the dpHNL and LSM-mediated models, all of which yield decay signatures inside the detector of interest. The three production locations are
\begin{enumerate}
    \item Dump: production in the $4\times 4\times 4.21$ m$^3$ iron dump
    \item Dirt: production in the dirt upstream of the given detector
    \item Detector: production in the detector itself.
\end{enumerate}
We will see that there are complementary strengths and weaknesses in each of these production locations. For instance, production in the dirt benefits from a large number of targets for potential neutrino upscattering. In contrast, production in the dump will yield out-of-time signatures (relative to the incoming neutrino beam) which will allow for greater signal/background separation.

MiniBooNE has observed a significant excess of single-electron or single-photon-like events at low energies, reaching nearly $5\sigma$ significance~\cite{MiniBooNE:2020pnu}. These events could be explained by dpHNL events decaying into photons in the detector~\cite{Kamp:2022bpt, Vergani:2021tgc}. Ref.~\cite{Kamp:2022bpt} analyzed MiniBooNE, MINERvA, and MicroBooNE data in this model, also allowing for $\nu_\mu \to \nu_e$ oscillations to contribute to the MiniBooNE low-energy excess via an eV-scale sterile neutrino. In doing so, the dpHNL can explain a significant fraction of the excess MiniBooNE events while remaining consistent with other experimental results. Two specific sets of parameters were explored in detail, corresponding to fits to MiniBooNE's excess in terms of its reconstructed neutrino energy or the shower direction, yielding
\begin{align}
    (m_N, d_\mu) &= \begin{cases} \left(0.47\ \mathrm{GeV}, 1.25\times 10^{-6}\ \mathrm{GeV}^{-1}\right) & \mathrm{energy{-}spectrum\ fit} \\
    \left(0.08\ \mathrm{GeV},\ 1.7\times 10^{-7}\ \mathrm{GeV}^{-1}\right) & \mathrm{direction\ fit}
    \end{cases}.
\end{align}
We will use these as benchmark points to examine the observable features in the SBN detectors and to distinguish between different production locations of the dpHNL.

We estimate expected/known background rates for the signatures of interest in~\cref{tab:ExperimentalDetails} -- rates for MiniBooNE and MicroBooNE come from Refs.~\cite{MiniBooNE:2020pnu} and~\cite{Mogan:2021iau}, respectively (with POT-rescaling performed for MicroBooNE). Background rates for SBND and ICARUS are then extrapolated from the MicroBooNE $1\gamma0$p topology by scaling according to differences in neutrino flux, detector mass, and POT. We emphasize that this procedure provides only a rough estimate, as selection efficiencies like energy and angular cuts, etc. and detector-specific effects like detector efficiencies, etc. are not accounted for. A full simulation and analysis would be required for a more accurate prediction, but such a study is beyond the scope of this work. We include this approximation to facilitate a comparative context for the different detectors.

\subsection{Existing dpHNL Constraints}
The existing constraints on the dpHNL model considered here come from CHARM~\cite{CHARM:1988tlj, CHARM-II:1994dzw}, LSND~\cite{LSND:2001akn}, Borexino~\cite{Bellini:2011rx}, Super Kamiokande~\cite{Super-Kamiokande:2001ljr}, NOMAD~\cite{NOMAD:1998pxi, NOMAD:1997pcg, Vannucci:2014wna}, and SN~1987A~\cite{Chang:2016ntp, Kamiokande-II:1987idp, Fischer:2016cyd, Dreiner:2013mua, Dreiner:2003wh}. NOMAD searched for signals directly from the $N \rightarrow \gamma \nu_\mu$ channel~\cite{Gninenko:2012rw}. CHARM-II, meanwhile, constraints the $d_{\mu}$, by requiring that the neutrino transition magnetic moment cross-section be below the precision of their measurement of the neutrino-electron cross-section~\cite{Coloma:2017ppo}. LSND, a beam-dump experiment similar to MiniBooNE, places additional limits as calculated in Ref.~\cite{Magill:2018jla}. Constraints from SN 1987A arise from the supernova's cooling rate~\cite{Magill:2018jla}. Borexino sets limits based on the search of $N \rightarrow \nu + e^+ + e^-$, where $N$ is produced via nuclear beta decays in the Sun~\cite{Plows:2022gxc, Plestid:2020vqf}. Super-Kamiokande's limits stem from the absence of atmospheric neutrino upscattering to $N$ and its subsequent decay via the dipole portal~\cite{Plestid:2020vqf, Li:2024gbw, Gustafson:2022rsz}.

\section{Signal Distributions}\label{sec:Distributions}
Because of the different production locations,  dpHNL events originated in the dump, dirt, and detector carry significantly different event kinematics. In this section, we characterize those differences and discuss how they may be used to (a) leverage better signal-to-background separation, and (b) (in the case of a signal excess) discern between different production-method contributions.

In general, we focus on three main observables which can differ between the different distributions
\begin{enumerate}[label=(\alph*)]
\item the photon energy $E_\gamma$ observable in the event,
\item the direction of the photon with respect to the beam axis, $\cos\theta_\gamma$, and
\item the time of the event with respect to the expected SM neutrino beam spill.
\end{enumerate}
In general, the combined observation of these three observables will provide excellent separation between production in each of the three locations, as well as separation from SM neutrino background events.

\begin{figure}[!htbp]
    \centering
        \includegraphics[width=\textwidth]{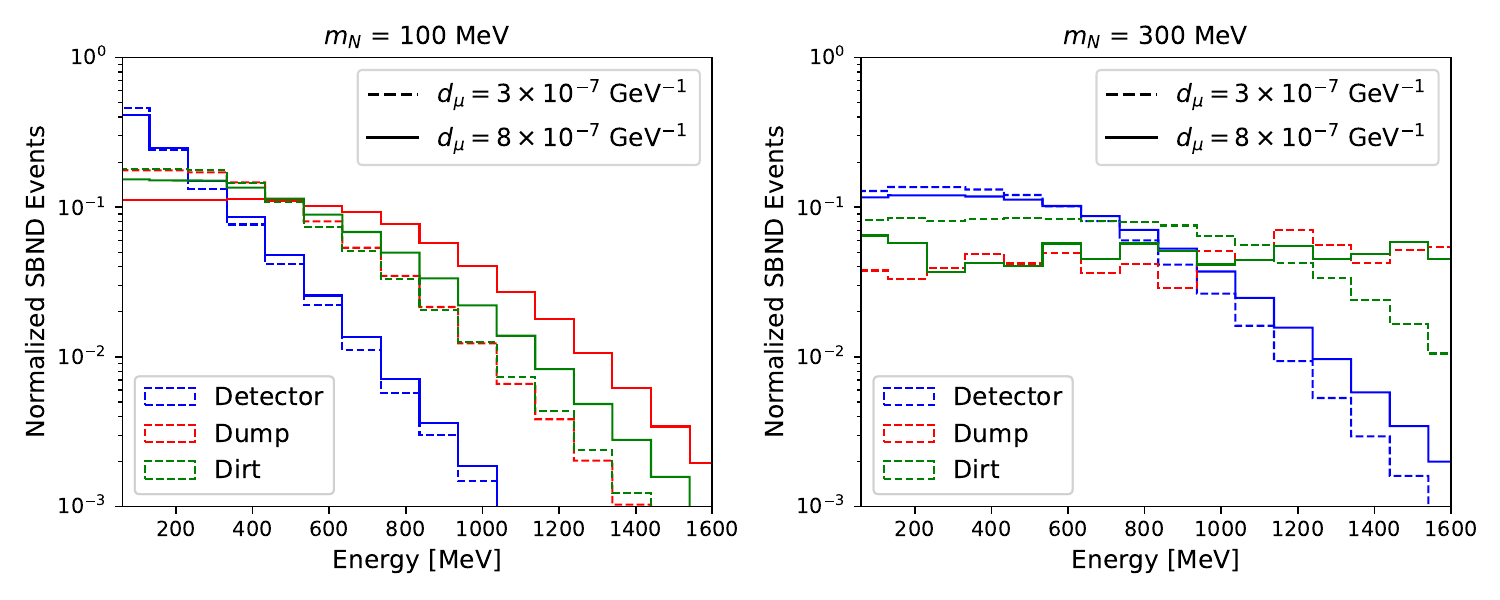}
    \captionsetup{justification=Justified, singlelinecheck=false}
    \caption{Energy spectra of outgoing photons produced by the decay of dpHNLs in the SBND detector. Each colored line corresponds to a different dpHNL production location. The x-axis represents the energy of the outgoing photon in the lab frame, while the y-axis shows the normalized number of SBND events per energy bin. Dashed and solid lines indicate values of $d_\mu=3 \times 10^{-7}$~GeV$^{-1}$ and $d_\mu=8 \times 10^{-7}$~GeV$^{-1}$, respectively. The left and right plots correspond to dpHNL masses of $m_N=100$~MeV and $300$~MeV, respectively.}
    \label{fig:EnergySpectrum}
\end{figure}

\paragraph{Energy spectra -- }\label{sub:energy} Characteristic energy spectra are shown for several parameter points of interest in~\cref{fig:EnergySpectrum}. We focus on two masses, $m_N = 100$~MeV and $300$~MeV, representative of the range of interest for the SBN detectors. For specificity, we focus on event distributions in SBND, although the other SBN detectors would observe qualitatively similar behavior.
In order to emphasize the differences between the three production locations, we area-normalize these expected signal distributions. We also consider multiple values of the dipole constant $d_\mu$, $3\times 10^{-7}$~GeV$^{-1}$ and $8\times 10^{-7}$~GeV$^{-1}$, which effectively determine whether dpHNL is long- or short-lived.

Generically, we find that the events arising from dpHNL produced in the SBND detector itself lead to the lowest-energy signals. This comes from the requirement of a short lab-frame decay length in order for such signals to be visible. Signatures from dpHNLs produced in the dirt upstream of the
detector, or even more prominently in the BNB iron dump, lead
to higher-energy signatures. This is for multiple reasons: first, those particles will benefit from a larger boost factor to survive until the SBND detector where they decay. Secondly, dpHNL with higher energies will typically be directed more in the forward-going region because of the nature of the Primakoff upscattering process. As we vary $m_N$ and $d_\mu$, these qualitative takeaways hold and some of the differences are emphasized as the dpHNL lifetime becomes relevant for the distances of interest in this experimental setup.

\begin{figure}[!htbp]
    \centering
    \includegraphics[width=\textwidth]{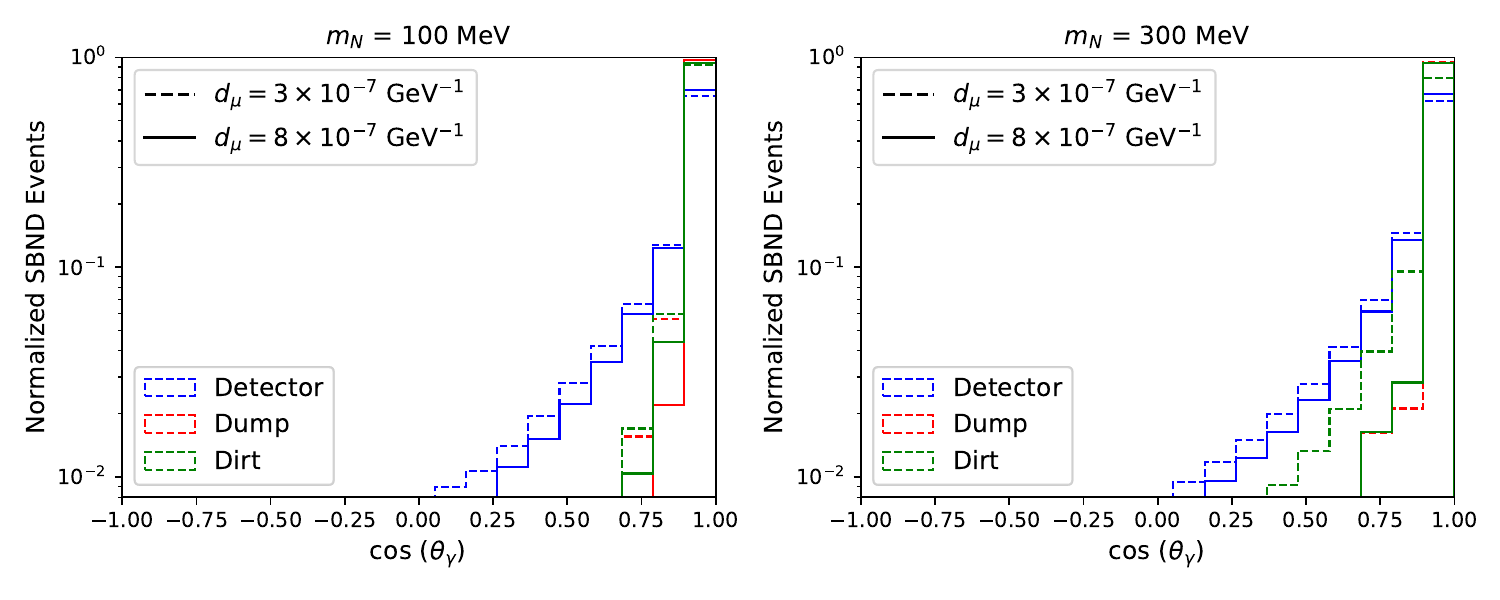}
    \captionsetup{justification=Justified, singlelinecheck=false}
    \caption{Angular spectra of outgoing photons produced by the decay of dpHNLs in the SBND detector. Each colored line corresponds to a different dpHNL production location. The angle $\theta_\gamma$ is defined by the outgoing photon direction relative to the beam axis. Dashed and solid lines indicate values of $d_\mu=3 \times 10^{-7}$~GeV$^{-1}$ and $d_\mu=8 \times 10^{-7}$~GeV$^{-1}$, respectively. The left and right plots correspond to dpHNL masses of $m_N=100$~MeV and $300$~MeV, respectively.}
    \label{fig:AngularSpectrum}
\end{figure}

\paragraph{Angular spectra -- }\label{sub:angular} Another observable that is useful for distinguishing these signal models (as well as distinguishing all signal models vs.~those from neutrino-scattering backgrounds) is the direction of the outgoing photon. The separation of different signals' production locations is evident in~\cref{fig:AngularSpectrum}, where we choose the same combinations of $m_N$ and $d_\mu$ as when exploring the energy spectra in~\cref{fig:EnergySpectrum}.

For events originating in the detector, we see that the spectrum peaks towards forward-going events, however, there is significant support down nearly to $\cos\theta_\gamma = 0$. It is this exact feature that provides evidence for the dpHNL, over other long-lived particle models, to explain the MiniBooNE low-energy excess~\cite{MiniBooNE:2018esg, MiniBooNE:2020pnu} -- long-lived particles produced far upstream of the detector will lead to $\cos\theta_\gamma \approx 1$, where the MB low-energy excess extends to relatively small values of this observable. 

We see this behavior evident in the spectra of events where the dpHNL was produced in the dirt and dump, both of which peak even more significantly at $\cos\theta_\gamma \to 1$ -- this is because the dpHNL must be very forward-going (and high-energy, as we witnessed in~\cref{fig:EnergySpectrum}) in order to reach and decay inside the detector. This will naturally lead to relatively high-energy, forward-going photons in SBND. Overall, we see that these features are not too sensitive to the choice of $m_N$ or $d_\mu$.

\begin{figure}[!htbp]
    \centering
    \begin{subfigure}{0.48\textwidth} 
        \centering
        \includegraphics[scale=0.57]{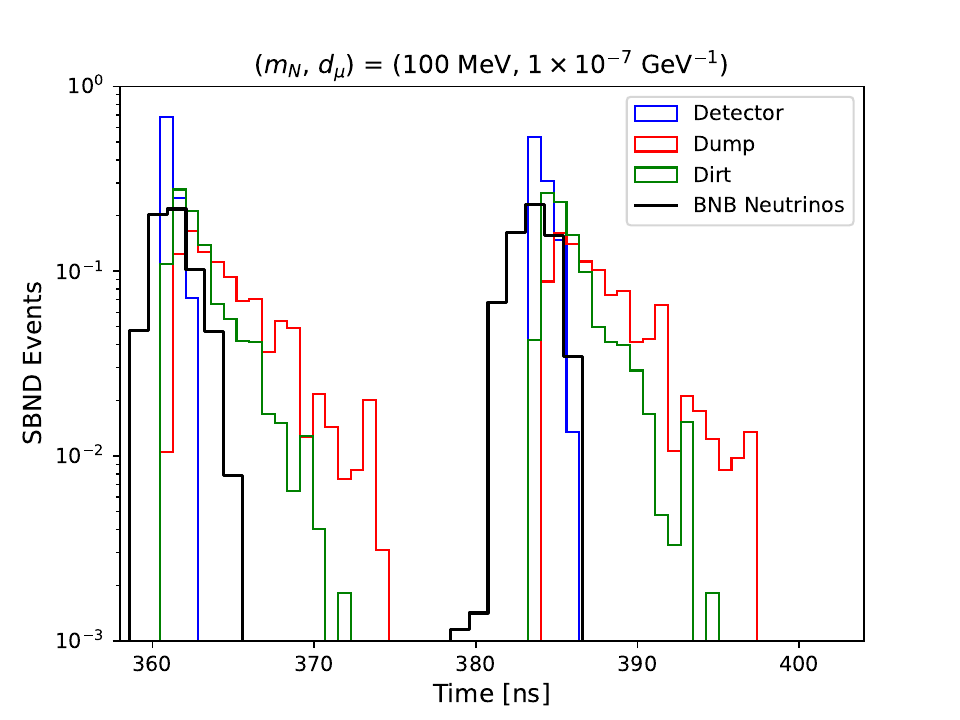}
    \end{subfigure}
    \begin{subfigure}{0.48\textwidth} 
        \centering
        \includegraphics[scale=0.57]{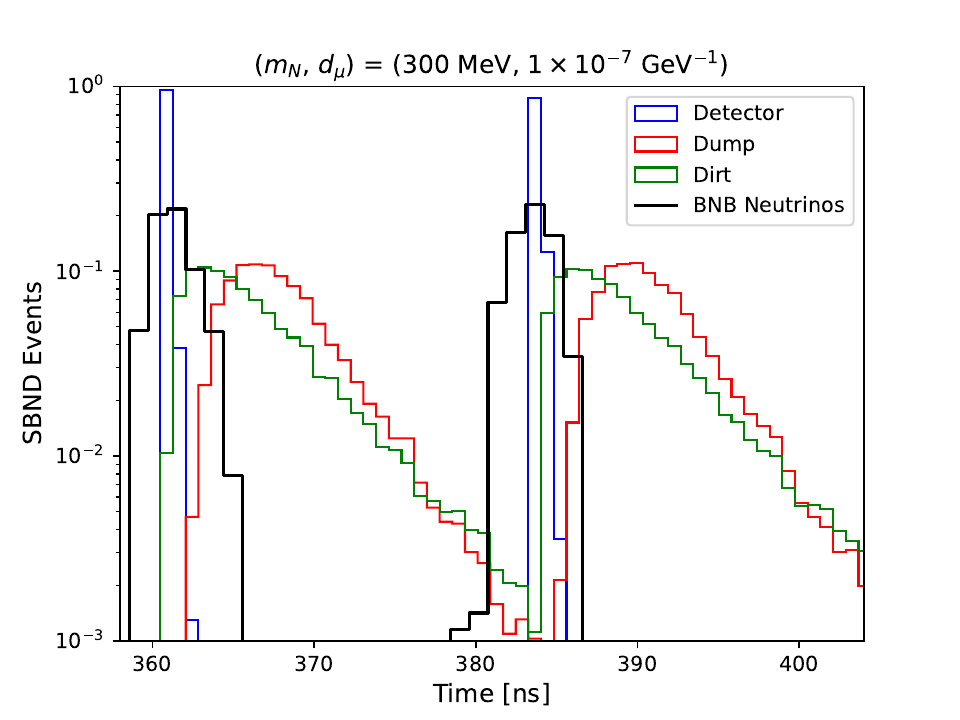}
    \end{subfigure}
    \captionsetup{justification=Justified, singlelinecheck=false}   
    \caption{Timing spectra of dpHNLs reaching the front face of the SBND detector for the dump and dirt lines, and $\nu_\mu$ for the detector lines. Here, $t=0$ represents the moment when the charged mesons cross the end of the magnetic horn. Only dpHNLs that produce visible signals in the detector are considered. The time on the x-axis represents the total time from $t=0$ until the dpHNLs/$\nu_\mu$ reach the front face of the SBND detector. The black line shows the timing spectra from the SBND simulation of the BNB $\nu_\mu$ flux reaching the detector.~\cite{SBND:2024vgn, SearchesforBSM, SearchesforBSM1}.}
    \label{fig:TimingSpectrum}
\end{figure}

\paragraph{Timing spectra -- } The final observable, with particular weight in separating signal from background, is the time of the decay event occurring with respect to the proton beam spill. We display this in~\cref{fig:TimingSpectrum} 
for two masses of interest. For comparison, we also display the expected time-structure of neutrino events from the BNB $\nu_\mu$ flux~\cite{SBND:2024vgn, SearchesforBSM, SearchesforBSM1}, where further discussions and improvements have demonstrated that nanosecond-level event identification is possible in SBND.

In these distributions, $t=0$ corresponds to the time at which SM mesons are produced at the BNB target; the pulsed structure of the protons striking the target leads to the periodic motion (${\sim}$18~ns) present throughout the timing distributions. Background events resulting from the BNB neutrinos arrive first, as the SM neutrinos travel effectively at light speed. For the same reason, dpHNL events in which the upscattering occurs in the detector are early, whereas those originating from the dirt and the dump experience time delays due to the lower velocities of the heavy $N$. The arrival delay becomes even more pronounced for heavier $m_N$, apparent in~\cref{fig:TimingSpectrum} (right), where the separation of the different signal curves becomes clearer.

To enable a comparison between the spectra at the nearest detector (SBND) and the furthest detector (ICARUS), the corresponding energy, angular, and timing spectra for ICARUS are provided in~\cref{app:MoreComparisons}.

\subsection{dpHNL Model vs LSM Model}
In~\cref{subsec:LSMModel} we introduced an alternate model in which the upscattering into heavy fermions is mediated by a light scalar. This subsection explores the kinematic differences that exist, contrasting between the light-scalar- and photon-mediated cases. The key difference lies in the massive mediator, which causes the neutrino-to-$N$ upscattering to have markedly different energy and angular dependence.

\begin{figure}[!htbp]
    \centering
    \begin{subfigure}{1.0\textwidth} 
        \centering
        \begin{subfigure}{0.48\textwidth}
            \centering
            \includegraphics[scale=0.50]{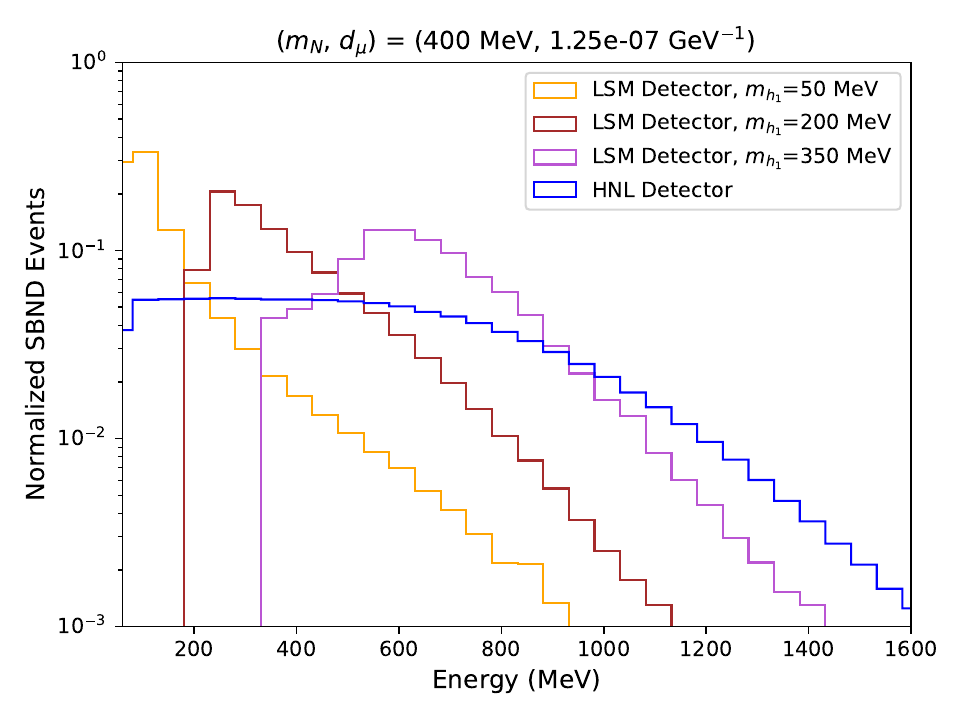}
        \end{subfigure}
        \begin{subfigure}{0.48\textwidth}
            \centering
            \includegraphics[scale=0.50]{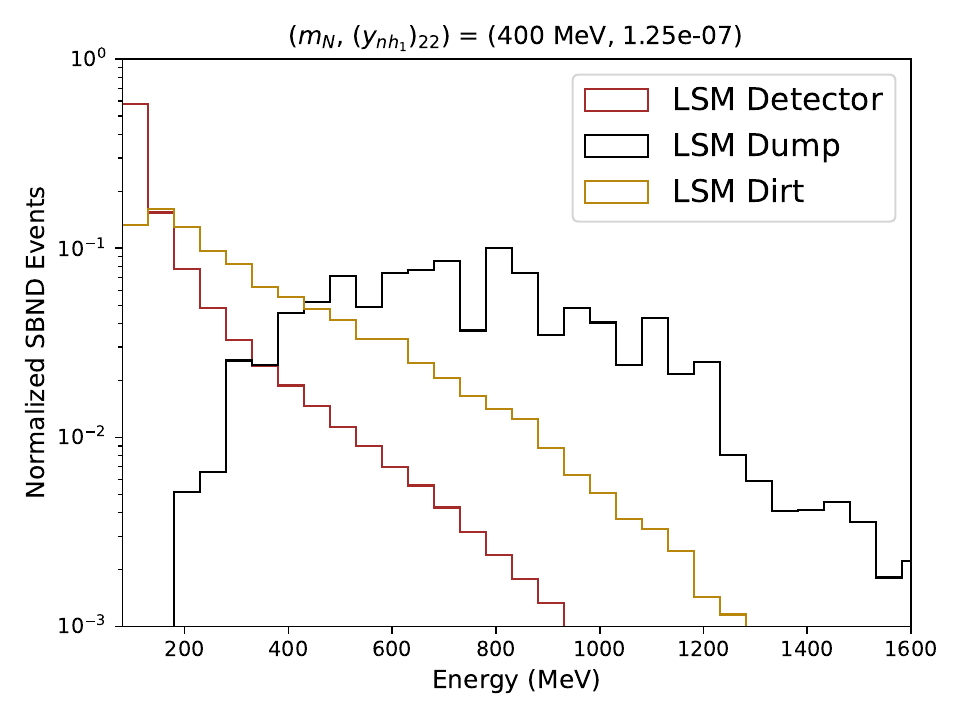}
        \end{subfigure}
    \captionsetup{justification=Justified,singlelinecheck=false}
        \caption{Energy Spectra of Outgoing Photons and $h_1$ produced from decaying of HNLs in the HNL and LSM models respectively.}
    \end{subfigure}
    \begin{subfigure}{1.0\textwidth} 
        \centering
        \begin{subfigure}{0.48\textwidth}
            \centering
            \includegraphics[scale=0.50]{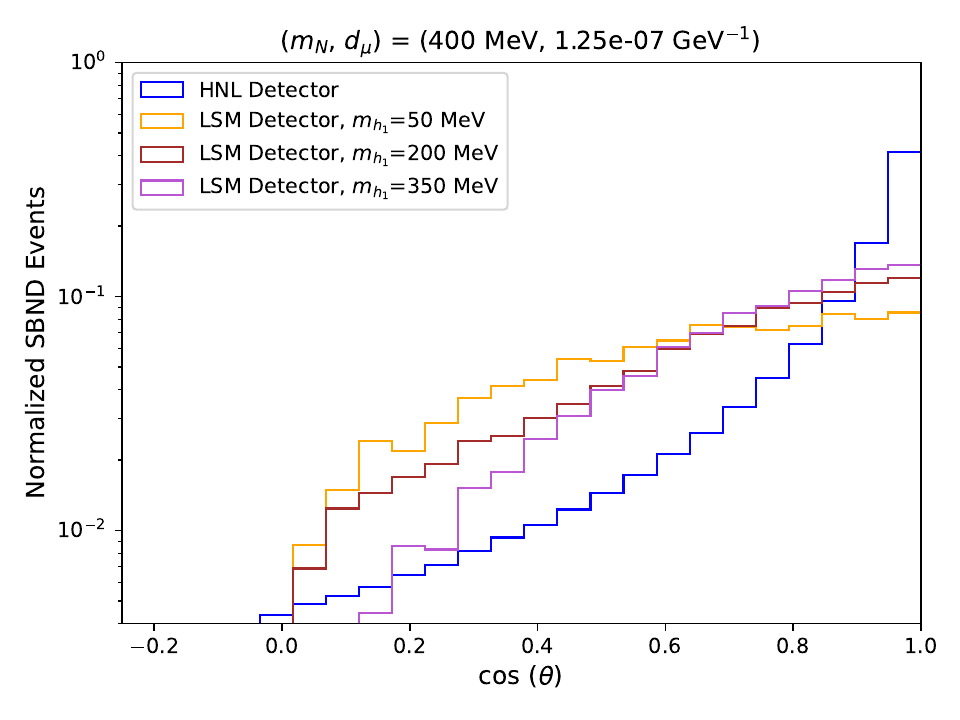}
        \end{subfigure}
        \begin{subfigure}{0.48\textwidth}
            \centering
            \includegraphics[scale=0.50]{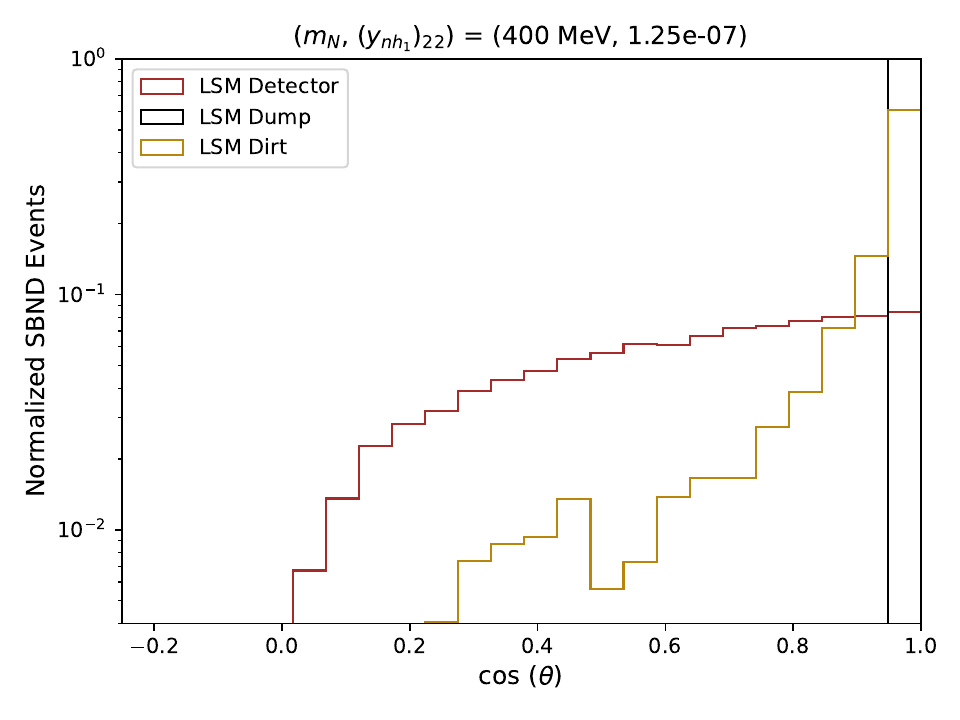}
        \end{subfigure}
    \captionsetup{justification=Justified,singlelinecheck=false}
        \caption{Angular Spectra of Outgoing Photons and $h_1$ produced from decaying of HNLs in the HNL and LSM models respectively.}
    \end{subfigure}
    \captionsetup{justification=Justified,singlelinecheck=false}     
    \caption{Energy and Angular Spectra of the outgoing photons~($h_1$) produced from decaying of the HNLs in the SBND detector for the dpHNL~(LSM) model. Each of those three lines denotes where the HNL has been produced. The energy on the x-axis represents the energy of the outgoing photon~($h_1$) in the lab frame. The $\cos \theta$ on the x-axis of the angular spectra is the cosine of the angle that the outgoing photon~($h_1$) makes with the beam-axis for the dpHNL~(LSM) model. Parameter values considered for the LSM model are as follows: $m_{h_1}=80$ MeV, $(y_{nh_1})_{22}=d$ (without units), $y_{uh_1}=y_{dh_1}=5\times 10^{-3}$, $y_{eh1}=1 \times 10^{-7}$.}
    \label{fig:dpHNL_LSM_AngularEnergyComp}
\end{figure}

First,~\cref{fig:dpHNL_LSM_AngularEnergyComp} demonstrates the difference in energy and angular spectra between the two signal models. For concreteness, among the different production points' comparisons in the LSM model (right plots of~\cref{fig:dpHNL_LSM_AngularEnergyComp}), we fix $m_{h_1} = 80$~MeV~\cite{Dutta:2020scq}. When comparing energy-related observables, the energy corresponds to the total visible energy in the decay, i.e.\ the photon in the dpHNL decay ($N\to \nu\gamma$) and the electron/positron pair in the light-scalar mediated case ($N \to \nu h_1,\ h_1 \to e^+ e^-$).

Similar to the dpHNL case, this light-scalar mediated scenario prefers higher-energy events when the origin is in the BNB dump, for very similar kinematical reasons. The same is true of the distribution of event directions, as events with $N$ arriving from the dump will tend to be very forward-going in order to reach and decay inside the detector. Events originating inside the detector follow a characteristically different spectrum -- due to the large $m_{h_1}$, the eventual $e^+ e^-$ showers will be generated at wider angles with respect to the beam than in the dpHNL case.

When we compare the flux between the two models, we observe that the dpHNL model results in higher visible energies as compared to the LSM model for the same production region. This is attributed to the choice of the scalar's mass and coupling. Since the scalar is massive, the neutrino upscattering in the LSM model has more energy transfer to the nucleus as compared to the photon exchange in the dpHNL model. This results in softer LSM HNLs, and hence softer visible energy spectra. The coupling of the the scalar $h_1$ to $e^+ e^-$ determines the lifetime of the scalar and therefore, its probability to decay within the detector. This dependence comes into the picture especially while comparing detector fluxes where the choice of low coupling $y_{eh_1}$ causes the high momentum scalars to escape from the detector. Hence, the detector flux in the LSM model is more peaked at low energies than the dpHNL model. This pattern is consistently observed across all masses.

We note that in the LSM model, the signal comes not from $h_1$ itself but from its subsequent decay into an electron-positron pair. As a result, $h_1$ can be produced either inside or outside the detector and must be sufficiently forward-boosted to reach the detector. In contrast, the dpHNL model requires the dpHNL to reach the detector, allowing for less stringent boosting requirements compared to the LSM model. Therefore, in the dpHNL model, the photon produced need not be as strongly forward-boosted for the dirt and dump contributions.

The sudden spikes in the energy and angular plots are attributed to noise, as there are only a few events at those specific mass and coupling points. We also examined the angular spectra of the opening angles between $e^{+}$ and $e^{-}$ in the LSM model. An analysis of the timing spectra for the LSM model revealed no differences compared to the dpHNL model, with similar features observed in both cases. Since SBND can utilize the PRISM strategy~\cite{PRISM}, various other comparisons can be conducted— such as spatial distributions across the detector area, events vs. radius plots, etc. However, we notice that the most prominent features can be observed in the angular and energy spectra visualization that are presented in Fig.~\ref{fig:dpHNL_LSM_AngularEnergyComp}.

\section{Signal-vs-Background Separation}\label{sec:SignalVsBackground}
Throughout~\cref{sec:Distributions}, we discussed distributions of several observables of interest for the dpHNL -- similar arguments can be made for other classes of models that rely on upscattering and decay, such as the light-scalar mediator model. In this section, we change our emphasis to that of signal/background separation and leverage these distributions to best enable such discrimination.

In~\cref{tab:ExperimentalDetails}, we provided background rates for single-photon searches in MiniBooNE~\cite{MiniBooNE:2020pnu} and MicroBooNE~\cite{Mogan:2021iau} based on existing searches, as well as estimates for such searches in SBND and ICARUS (relying on exposure/detector-mass rescaling arguments). This assumes that backgrounds are dominated by neutrino scattering -- a reasonable assumption -- where neutral-current-single-pion (NC$\pi^0$) scattering dominates.

When discussing the timing spectra of signal events surrounding~\cref{fig:TimingSpectrum}, we compared these distributions against expected BNB neutrino (background) events, observing the key differences between the BNB background distribution and, especially, signals originating in the dirt upstream of SBND as well as in the BNB dump. If one designed a specific search for dpHNL originating in the BNB dump, then a strong timing cut (to greatly reduce backgrounds while retaining the bulk of the signal) can be placed. Thus, it is natural from timing distributions alone to expect a lower background rate for dump-originated dpHNL than in searches for events originating in the detector or the dirt.

Similar arguments for background reduction can be made by leveraging the energy (\cref{fig:EnergySpectrum}) and angular (\cref{fig:AngularSpectrum}) distributions of the different classes of signal events. Background events from NC$\pi^0$ originate from misidentification in SBND, where the $\gamma\gamma$ pair is either overlapping or one of the photons is too low energy to be reconstructed, for instance. Such events will tend to populate lower energies (${\sim}$100s of MeV) and be relatively isotropic due to the production of $\pi^0$ in neutrino scattering and the subsequent $\pi^0 \to \gamma\gamma$ decay. Given the distributions of~\cref{fig:EnergySpectrum,fig:AngularSpectrum}, we expect that this will favor the dpHNL produced by dump and dirt to separate the signal from the background.

In the analyses to follow, we will not attempt to quantify a possible background reduction due to these kinematics, nor the relative strengths of background reduction for the different signal classes. Nevertheless, we hope that highlighting these kinematical distributions will lead to more dedicated experimental analyses aimed at searching for all different production mechanisms and locations.

\section{Model Sensitivity \& Discussion}\label{sec:ResultsDiscussion}
As discussed in~\cref{sec:SignalVsBackground}, here we do not attempt to reduce expected background rates by leveraging event kinematics, although we expect in doing so that we could perform searches for dirt and dump-produced dpHNL with fewer expected background events. Instead, we conservatively assume a constant background rate (given for each experiment in~\cref{tab:ExperimentalDetails}) for each experiment. Even when only studying a specific dpHNL production location,
we consider the total background rate in determining sensitivity.

We construct a simple, statistics-limited $\chi^2$ function to determine sensitivity,
\begin{align}\label{eq:ChiSquared}
\chi^2 = \frac{s^2}{s + b},
\end{align}
where $s$ and $b$ are the expected signal and background rates, respectively, for a given analysis. To determine $90\%$~CL sensitivity, we require $\chi^2 > 4.61$.

First, we explore the parameter space associated with the dpHNL for the MiniBooNE and SBND experiments. We separate each experimental projection into three, searching explicitly for dpHNL events originating in the detector, the dirt, or the BNB dump. We note (see the discussion in~\cref{sec:SignalVsBackground}) that these analyses, especially the dump-oriented ones, should be taken as conservative as we have applied no background reduction throughout. Especially when we consider the combined analysis in the coming discussion, our results will represent very conservative projections.
\begin{figure}[!htbp]
    \centering
    \includegraphics[scale=0.45]{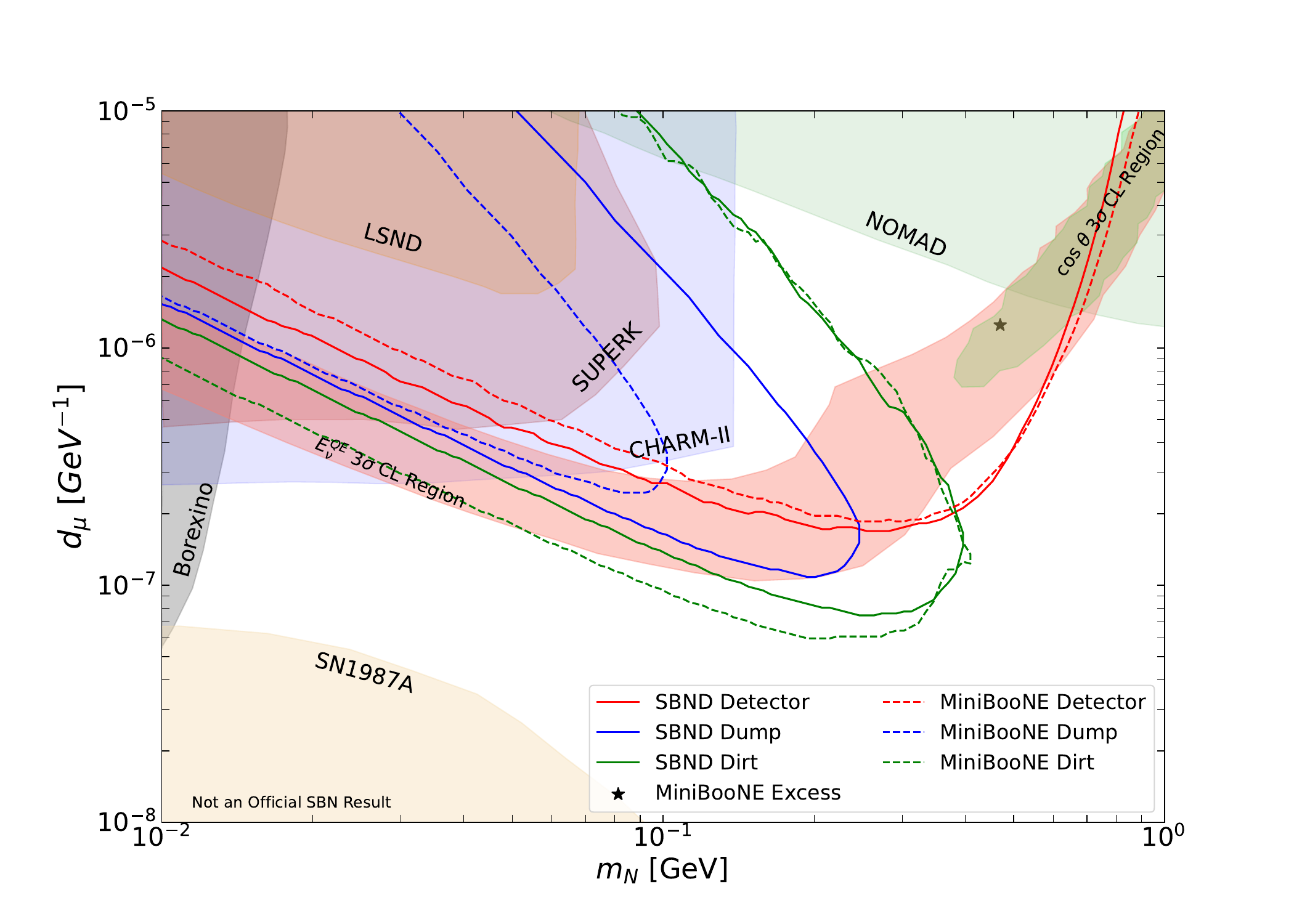}
    \captionsetup{justification=Justified,singlelinecheck=false}
    \caption{Potential (at 90\%~CL) for dipole-portal HNL discovery in SBND (solid lines) and MiniBooNE (dashed lines) as a function of mass and dipole strength. For each experiment, we divide the projected experimental capability into production in the given experiment's detector (red), in the dirt (green), or in the BNB dump (blue) before the dpHNL decays inside the detector. Comparisons against existing experimental constraints~\cite{Magill:2018jla} and potential parameter space to explain the MiniBooNE low-energy excess~\cite{Kamp:2022bpt} are included.}
    \label{fig:SBNDMiniBooNE_Separate}
\end{figure}

We display these projections for MiniBooNE and SBND in~\cref{fig:SBNDMiniBooNE_Separate}, where the solid (dashed) lines correspond to our expectations from SBND (MiniBooNE). We also compare against preferred regions of parameter space in the dpHNL scenario determined by Ref.~\cite{Vergani:2021tgc} (shaded red/green regions) in explaining the MiniBooNE low-energy excess. For each of the experiments, we show the complementarity of dpHNL production in the dump (blue), dirt (green), and detector (red) -- since the dpHNL produced in the detector can still be detected if they decay instantly, these contours reach the largest masses and do not have a sensitivity ``ceiling'' with respect to $d_\mu$, but those produced in the dump and dirt must be boosted enough to reach the detector and decay within. For SBND, we see that production in the dirt and dump are complementary, especially at masses $m_N \approx 100$~MeV and below.

In comparison to the detector and the dump, for both experiments, we highlight the power of scattering in the dirt -- the larger number of target nuclei and the relatively large-$Z$ silicon in the dirt allows for very efficient Primakoff upscattering of neutrinos into $N$ along the path of propagation.

As a final note regarding~\cref{fig:SBNDMiniBooNE_Separate}, we see that if dpHNL events are completely or partially responsible for the MiniBooNE low-energy excess, then a combination of analyses in SBND regarding dump-, dirt-, and detector-originating dpHNL will thoroughly test this hypothesis.

In~\cref{fig:SBNDMiniBooNE_Separate}, we focused on SBND and MiniBooNE, where we also see that due to MiniBooNE's larger distance (from the BNB target and dump), it is less sensitive to dpHNL produced in the dump. This becomes more pronounced at even larger distances, for instance at ICARUS. For completeness, we show the ICARUS and MicroBooNE sensitivity to dpHNL from these three different production locations in~\cref{app:MoreSensitivity}.

\begin{figure}[!htbp]
    \centering
    \includegraphics[scale=0.45]{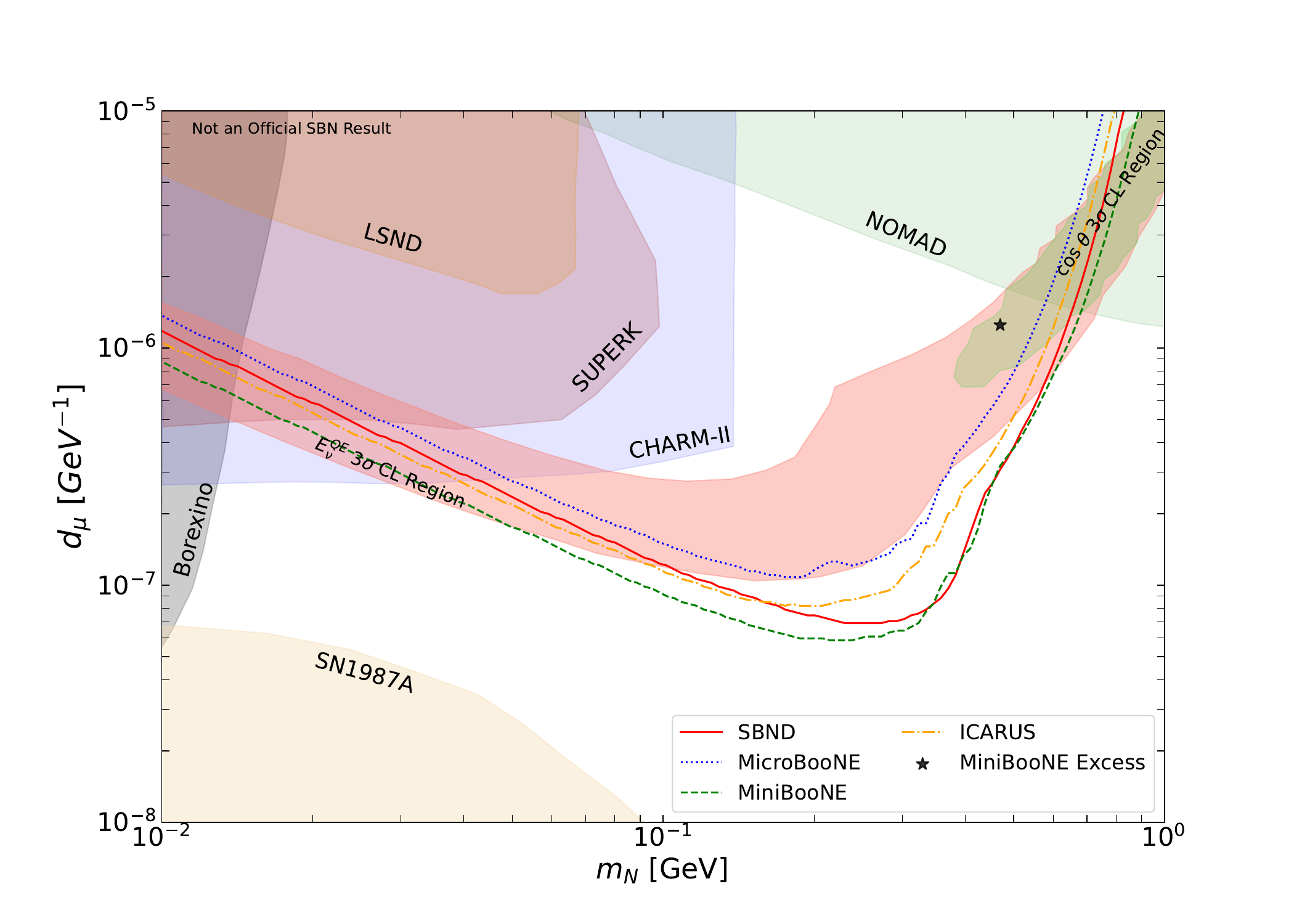}
    \caption{Similar to~\cref{fig:SBNDMiniBooNE_Separate}, considering production across all locations, for SBND (red), MicroBooNE (dotted blue), MiniBooNE (dashed green), and ICARUS (dot-dashed gold). All projections shown correspond to 90\%~CL.} 
    \label{fig:dpHNLTotal}
\end{figure}

A summary of the experimental analyses using all dpHNL origins combined is shown in~\cref{fig:dpHNLTotal}.
In determining these projections, we assume a total background (for each experiment, given in~\cref{tab:ExperimentalDetails}) compared against a combined signal, summing over dpHNL origins. Due to this construction and the potentially strong signal-vs.-background separation possible for dump-produced dpHNL, the projections should be taken as conservative. In~\cref{app:MoreSensitivity}, we provide additional comparisons, in the case of our SBND analysis, with different total background assumptions. Generically, we find that with a background of~$\sim 1000$, SBND offers the greatest sensitivity to dpHNL among LArTPC experiments due to its proximity to the BNB, as well as its capability to better leverage dirt- and dump-produced dpHNL.

\section{Summary and Conclusions}\label{SectionV}

In this work, we have incorporated contributions from neutrino upscattering at the iron dump showing that, particularly for close detectors like SBND, it plays a role as significant as the detector or dirt in determining the sensitivity of HNLs with transition magnetic moment.  Including the dump not only modifies the sensitivity but also reveals distinct features based on the dpHNL production point. Signals from the dump are the most forward-directed, followed by those from the dirt and then the detector. Furthermore, higher-energy photons detected in the detector are predominantly produced by dpHNLs originating from the dump. Our study clearly illustrates the varying energy and angular spectra at SBND, comparing characteristics across different masses and couplings. Furthermore, the timing spectra analysis highlights a clear distinction between the origins of dpHNLs and BNB neutrinos, with heavier dpHNLs showing a broader spread and a longer time delay. 

For distant detectors like MicroBooNE, MiniBooNE, and ICARUS, the dirt's contribution becomes increasingly important for masses ${\sim}250$~MeV. Across all experiments, the detector contribution becomes dominant for dpHNL masses above $250$~MeV.

We also compared the performance of all SBN programs and MiniBooNE, assessing both total and individual contributions. With approximately $\mathcal{O}$($10^4$) background events for SBND, MiniBooNE was found to probe the most extensive parameter space. While this estimated SBND background estimate is conservative, we show total sensitivity lines for lower background counts as well. By applying a timing cut, the SBND background can be significantly reduced, potentially enhancing its sensitivity beyond MiniBooNE. We save the detailed background analysis for the future, where all the above features need to be taken into account. Here, we anticipate that each production point will complement the other.

Additionally, we explored the specific features SBND would observe when investigating the parameter space necessary to explain the MiniBooNE anomaly, identifying several distinctive signals.

Finally, we examined the LSM model and identified unique features in the energy and angular spectra that distinguish it from the dpHNL model with transition magnetic moment. This analysis offers a framework that can be extended to other models, allowing for the exploration of various scenarios and a deeper understanding of different theoretical possibilities.

\section*{Acknowledgements}

We thank Wooyoung Jang for his work on \texttt{GEANT4} simulations. We also thank Matheus Hostert, Vishvas Pandey, Nick W. Kamp, Ryan Plestid, and Austin Schneider for their useful discussions.
This work of BD, DG, AK, and KJK is supported by the U.S. Department of Energy Grant DE-SC0010813.

\appendix

\section{Flux Construction and Simulation}\label{AppendixA}
\begin{figure}[!htbp]
\centering
\includegraphics[scale=0.35]{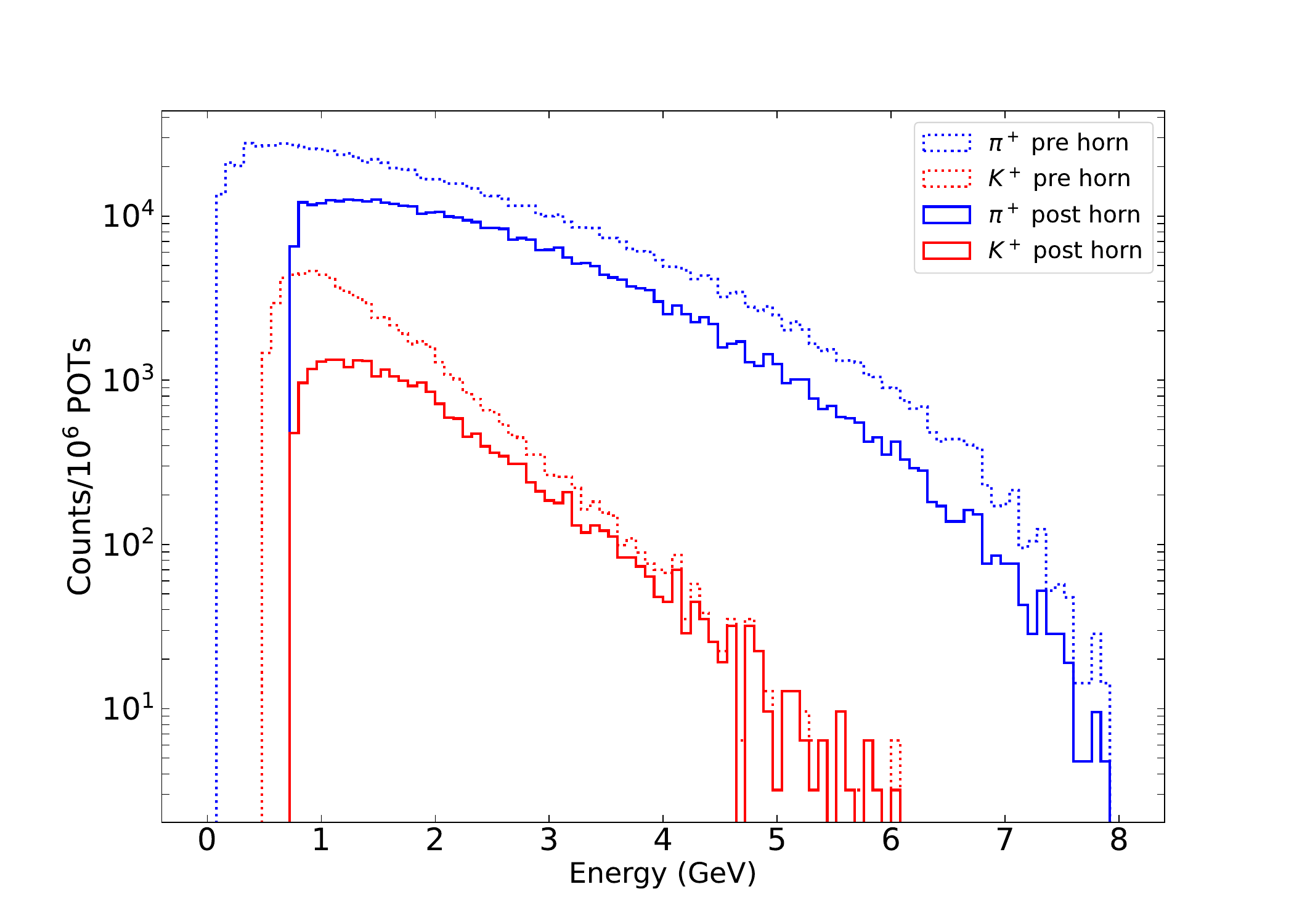}
\captionsetup{justification=Justified,singlelinecheck=false}
\caption{$\pi^+$ and $K^+$ fluxes pre and post-horn have been shown. The post-horn fluxes are obtained after imposing the angle and energy cuts. Comparison of the meson fluxes obtained before (dotted) and after (solid) the energy and angular cuts to replicate the magnetic horn effect.} 
\label{fig:1.7}
\end{figure}

In this section, we describe the fluxes used and their construction. The $\pi^{+}$ and $K^{+}$ fluxes were generated using \texttt{GEANT4}~\cite{Allison:2016lfl, GEANT4:2002zbu, Allison:2006ve} simulations. Since this analysis is for neutrino mode, where the POT is higher, we do not consider contributions from $\pi^-$ and $K^-$. To obtain the fluxes post the magnetic horn, we follow the prescription as detailed in Ref.~\cite{Dutta:2021cip} which was used to reproduce the MiniBooNE neutrino flux. In short, we select mesons with energy greater than 750~MeV and beamline angles between 0.03 and 0.2 radians to survive past the magnetic horn. We also assume that the mesons post-horn emerge parallel to the beamline ($p_T = 0$), and there are no meson decays inside the horn. ~\cref{fig:1.7} shows the meson flux before and after exiting the horn, based on the prescription described above.

After passing the horn, we collect the $\nu_\mu$ from the $\pi^+$ and $K^+$ that decay before the dump and make it to the SBND detector. We match this neutrino flux with the published SBND flux and find that our flux is~4.5 times larger. Therefore, we apply an overall, energy-independent weight of $1/4.5$ so that our total neutrino flux matches the published SBND ones.

\begin{figure}[!htbp]
\centering
\includegraphics[scale=0.35]{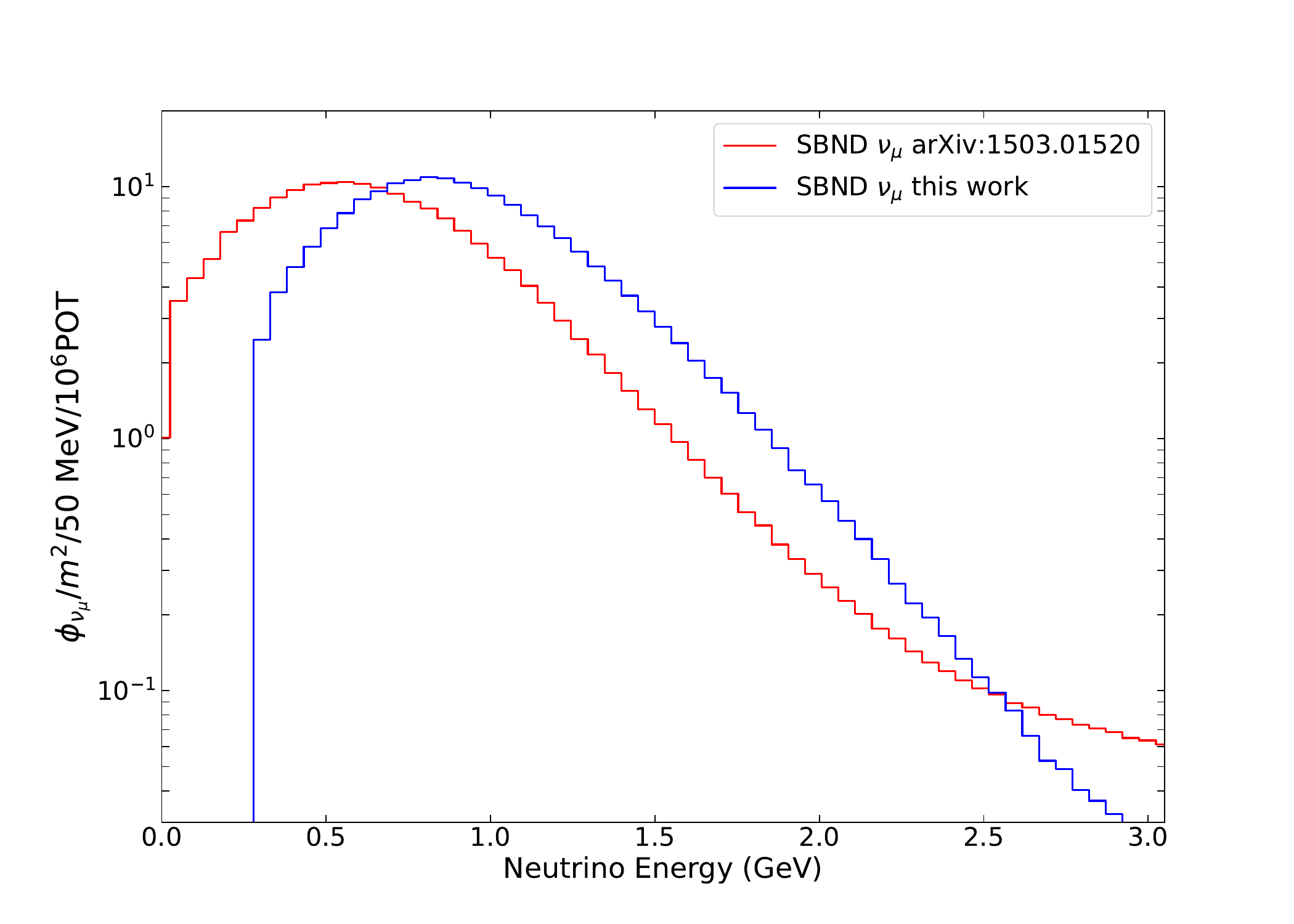}
\caption{The comparison between the published flux at the detector and the flux created for this work.} 
\label{fig:1.8}
\end{figure}

\begin{figure}[!htbp]
\centering
\includegraphics[scale=0.35]{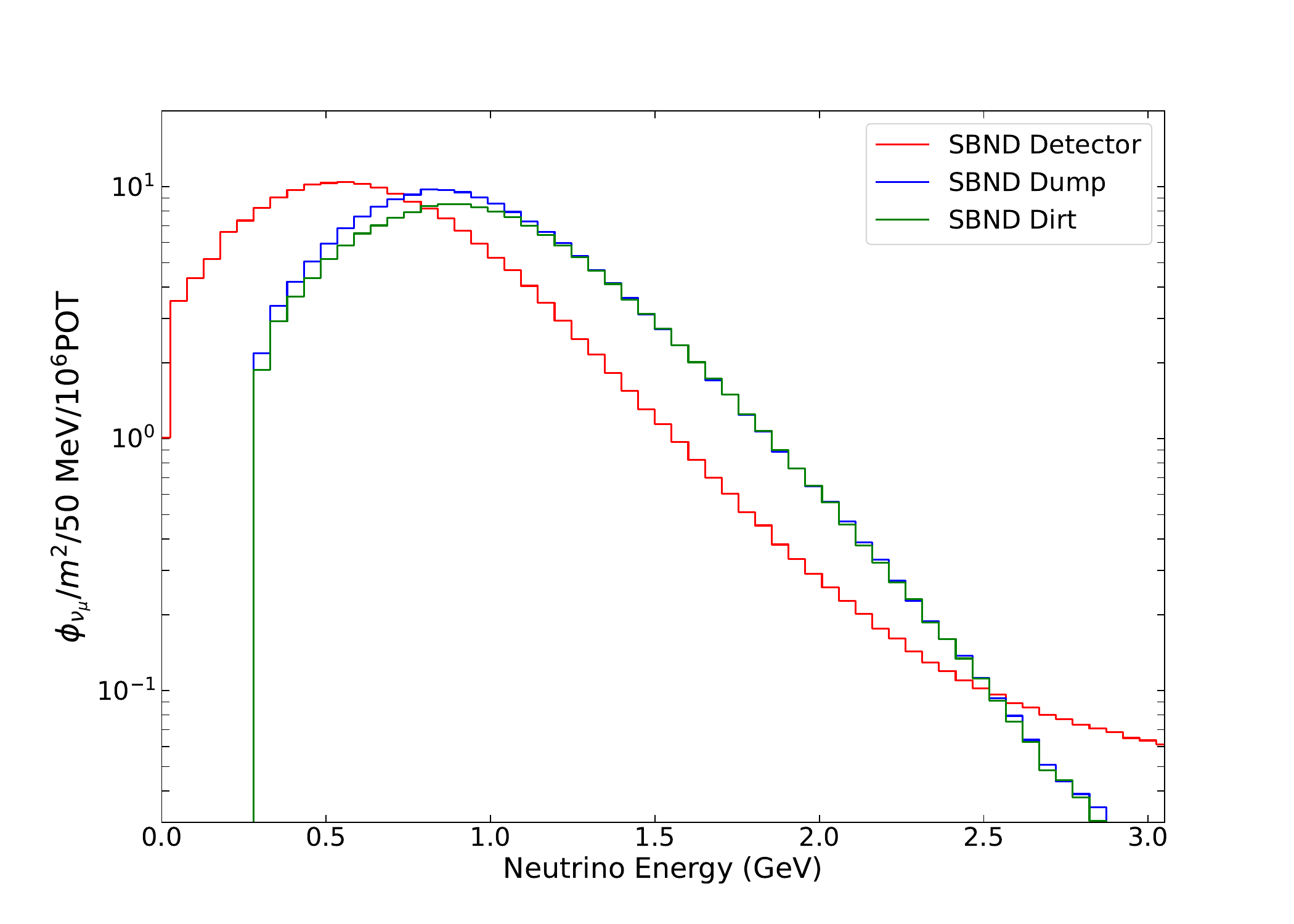}
\caption{Figure comparing the flux used in this analysis at three different locations -- the SBND detector (red, from official simulations), in the dirt upstream of SBND (green), and in the BNB dump (blue). Whereas the detector-location flux comes from official SBND fluxes, the latter two come from our independent simulations, described throughout~\cref{AppendixA}.} 
\label{fig:1.11a}
\end{figure}

While generating the upscattering of SM neutrinos at the dump, dirt, and detector, we include only those that enter and exit each of these entities. Since the detector is slightly off-axis, the above three contributions mildly differ from each other.~\cref{fig:1.8} compares the published and newly created $\nu_\mu$ fluxes at the detector. In reality, the horn geometry is significantly more complex. As a result of the approximations described above, differences arise between the published flux and the newly generated flux.~\cref{fig:1.11a} shows the three fluxes used in our analysis. Similar flux files were created for other SBN experiments as well.

\begin{figure}[!htbp]
\centering
\includegraphics[scale=0.45]{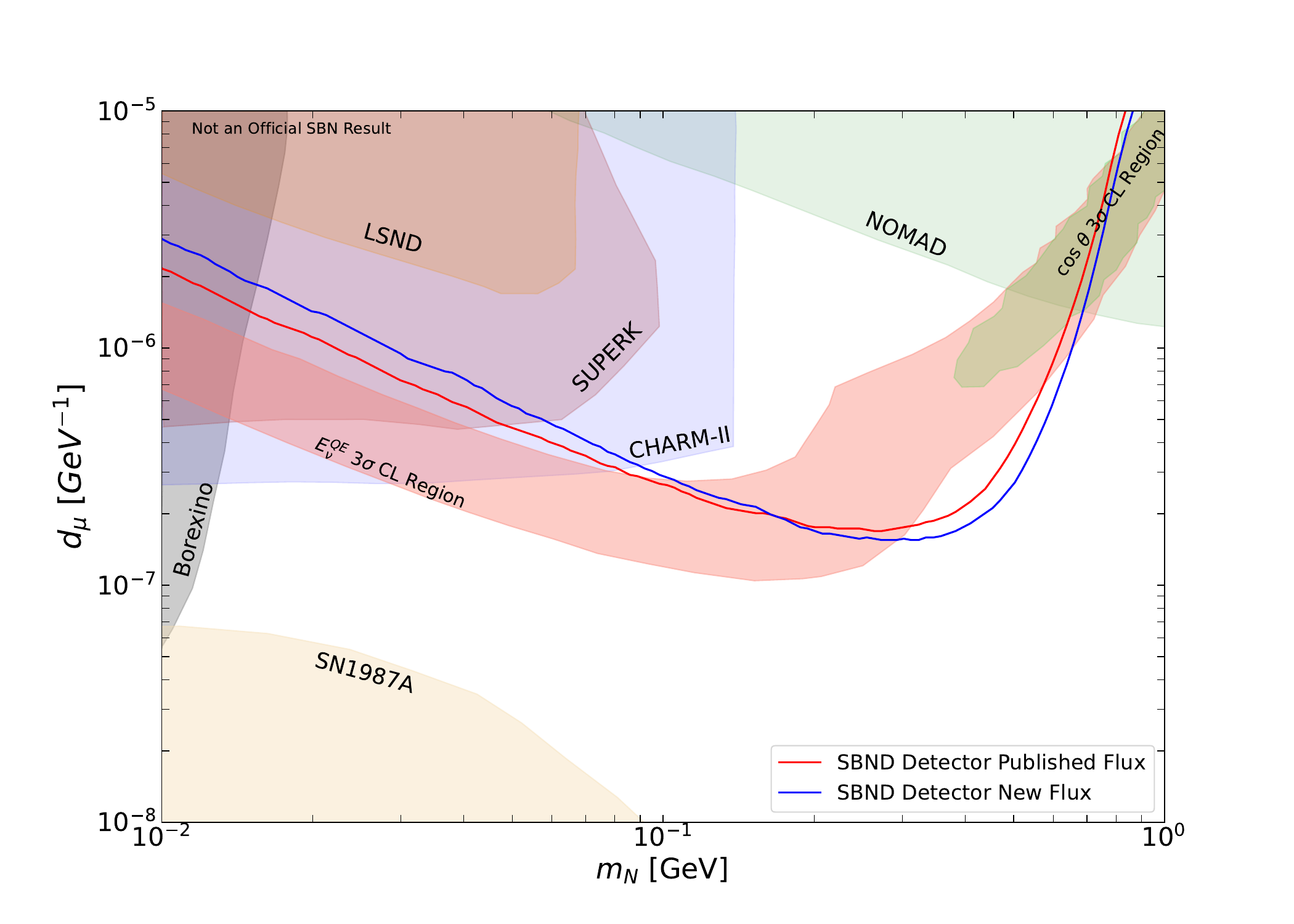}
\captionsetup{justification=Justified,singlelinecheck=false}
\caption{Sensitivity comparison of the dipole-portal heavy neutral lepton scenario considering production of dpHNL in the SBND detector volume -- the solid red (solid blue) line displays the sensitivity determined using the officially published (using our simulation-generated) $\nu_\mu$ flux at SBND.}
\label{FluxDiff}
\end{figure}

Finally, we wish to gauge the impact of our purpose-built simulation of neutrino fluxes for the analysis of the dipole-portal heavy neutral leptons (dpHNL). To do so, we re-calculate our expected sensitivity for the SBND experiment, considering production of dpHNL in the detector. For such a signal sample, we have our simulated flux (using the above-discussed technique) as well as the published SBND collaboration fluxes at the detector location for comparison. This juxtaposition is demonstrated in~\cref{FluxDiff}, where the red (blue) line displays the sensitivity using the published collaboration fluxes (using our simulated fluxes). At lower masses and couplings, the signal contribution from the published neutrino flux exceeds that of the flux simulated in this work. This occurs despite the simulated neutrino flux peaking at higher energies, as it decreases steeply at lower energies ($<$ 0.25 GeV), while the published flux retains a significant contribution in this region. Lower energetic dpHNLs are more likely to decay within the detector's length, further preferring the contribution from the $0-0.5$ GeV energy region. Conversely, at higher masses, the simulated flux dominates due to its peak being shifted to higher energies and its overall larger value at most energies up to approximately 2.5 GeV, beyond which both fluxes have negligible values. Comparing these, the difference in coupling for the same mass is at most a factor of 1.3, indicating a minimal impact on sensitivity to dpHNLs. Hence, we utilize the newly scaled neutrino flux, described above, to obtain the contribution for the dump and dirt, and the published neutrino flux for the detector for all the SBN programs.

\section{Kinematics}\label{AppendixB}
For completeness, we provide in this appendix some details of the scattering process of interest for the dipole-portal heavy neutral lepton (dpHNL) scenario we focus on for most of this work:
\begin{align}
    \nu~(p_1)\ +\ T~(p_2)\ \longrightarrow\ N~(p_3)\ +\ T~(p_4),
\end{align}
where the incoming neutrino $\nu$ scatters off a target nucleus $T$ (with mass $m_T$), producing the HNL $N$ with mass $m_N$, and four momenta as indicated. Specifically, we define some kinematical quantities for convenience: $q^\mu \equiv p_4^\mu - p_2^\mu = p_1^\mu - p_3^\mu$ is the momentum transferred to the nucleus, $E_\nu$ and $E_R$ are the incoming neutrino and outgoing nucleus energies, respectively, and $\theta_T$ ($\theta_N$) is the angle between the incoming neutrino and outgoing nucleus (dpHNL).

From these kinematics, we may determine the allowed range of $\theta_T$ given an incoming neutrino energy $E_\nu$. Generically, 
\begin{equation}\label{eq:thetaTRange}
\cos \theta_T = \frac{E_\nu E_R + m_T E_R + m_N^2/2}{E_\nu \sqrt{(m_T+E_R)^2 - m_T^2}}.
\end{equation}
By differentiation (solving $d\cos\theta_T/dE_R = 0$), we may obtain the maximum scattering angle $\theta_T$,
\begin{equation}\label{eq:thetaTmax}
\cos (\theta_T^{\rm max}) = \frac{m_N \sqrt{4m_T(E_\nu+m_T)-m_N^2}}{2m_T E_\nu}.
\end{equation}
Finally, by setting $\cos\theta_T^{\rm max} = 1$, we may obtain the allowed range of $E_R$ given an incident neutrino energy -- doing so results in the expression provided in~\cref{eq:ERMinMax}.

\section{Cross Section Calculations and Decay Widths}\label{AppendixC}

\subsection{dpHNL Model}
Referring to the Feynman diagram in Fig.~\ref{fig:upscatter}, the invariant matrix element of the Primakoff scattering process is given by the following expression:
\begin{widetext}
\begin{equation}\label{1.26}
|\mathcal{M}| = dZe\Bigg(\frac{1}{(p_2-p_4)^2}\Bigg)\Bigg(\bar{u}(p_3, m_N)\Bigg(\frac{1+\gamma_5}{2}\Bigg)((\slashed{p_2}-\slashed{p_4})\gamma^{\nu}-\gamma^{\nu}(\slashed{p_2}-\slashed{p_4}))u(p_1,0)\Bigg)\Bigg(\bar{u}(p_4,m_T)\gamma_\nu u(p_2,m_T)\Bigg)
\end{equation}
\end{widetext}

Squaring this matrix element, evaluating the traces in terms of Mandelstam variables using \texttt{FeynCalc}~\cite{Mertig:1990an, Shtabovenko:2016sxi}, and substituting it into $2\rightarrow 2$ scattering differential cross-section formula we obtain~\cref{eq:UpscatteringCrossSection}.

\subsection{LSM Model}

\begin{figure}[!htbp]
\centering
    \begin{subfigure}{0.35\textwidth}
        \begin{tikzpicture} 
            \begin{feynman} 
            \vertex(a0);\vertex[right=of a0](a1); \vertex[right=of a1](a2); \vertex[below=of a1](a3); \vertex[left=of a3](a4); \vertex[right=of a3](a5);
            \diagram*{ (a0)--[fermion, edge label'=\(\nu_{\mu}\)](a1),(a1)--[fermion, edge label'=\(N\)](a2),(a1)--[scalar, edge label=\(h_1\)](a3),(a4)--[fermion, edge label'=\(T\)](a3), (a3)--[fermion, edge label'=\(T\)](a5)}; 
            \end{feynman} 
        \end{tikzpicture}
    \caption{}
    \label{fig:LSMupscatter}
    \end{subfigure}
    \begin{subfigure}{0.35\textwidth}
        \begin{tikzpicture} 
            \begin{feynman} 
            \vertex(a0);\vertex[right=of a0](a1); \vertex[above right=of a1](a2); \vertex[below right=of a1](a3); 
            \diagram*{ (a0)--[fermion, edge label'=\(N\)](a1),(a1)--[fermion, edge label'=\(\nu_\mu\)](a2),(a1)--[scalar, edge label=\(h_1\)](a3)}; 
            \end{feynman} 
        \end{tikzpicture}
    \caption{}
    \label{fig:LSMdecay1}
    \end{subfigure}
    \begin{subfigure}{0.35\textwidth}
        \begin{tikzpicture} 
            \begin{feynman} 
            \vertex(a0);\vertex[right=of a0](a1); \vertex[above right=of a1](a2); \vertex[below right=of a1](a3); 
            \diagram*{ (a0)--[scalar, edge label'=\(h_1\)](a1),(a1)--[fermion, edge label'=\(l^+\)](a2),(a1)--[fermion, edge label=\(l^-\)](a3)}; 
            \end{feynman} 
        \end{tikzpicture}
    \caption{}
    \label{fig:LSMdecay2}
    \end{subfigure}
    \captionsetup{justification=Justified,singlelinecheck=false}
    \caption{Feynman diagram for (Top Left): Upscattering interaction mediated via light scalar $h_1$,  (Top Right): Decay of HNL into $\nu_\mu$ and $h_1$ in the LSM model and (Bottom): Decay of $h_1$ into $e^{+}$ and $e^{-}$ }
\end{figure}
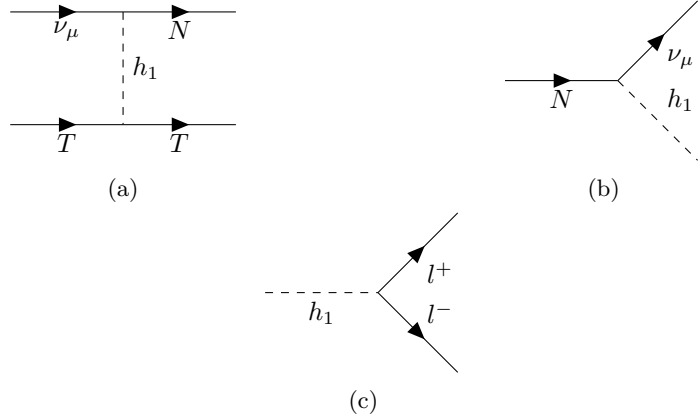

As described in Ref.~\cite{Dutta:2020scq}, this model is one of the simplest extensions of the scalar and fermionic sectors of the SM. In this model, the 2 Higgs doublet model with its complex scalar singlet extension is included, and the fermionic sector is extended with the addition of three right-handed sterile neutrinos to explain the neutrino masses via their mixing. The Lagrangian considered in this work is as follows:
\begin{equation}\label{1.28a}
\mathcal{L} \supset (y_{nh_1})_{22} h_1 \bar{N} \nu_\mu + (y_{eh_1})_{11}\bar{f}fh_{1} + \text{h.c.}
\end{equation}
Using the notations given in Ref.~\cite{Dutta:2020scq}, this Lagrangian allows an upscattering process mediated via light scalar $h_1$, as shown in Fig.~\ref{fig:LSMupscatter}. The HNL further decays to $\nu_\mu$ and $h_1$, followed by the decay of $h_1$ into $e^{+}$ and $e^{-}$. The kinematics of the upscattering process remain the same, although we cannot assume complete forwardness due to the massive nature of the mediator in the LSM model. 

The modified matrix element is as follows:
\begin{equation}\label{1.29}
|\mathcal{M}| = [\bar{u}(p_4)(Zf_p + (A-Z)f_n)_{11}u(p_2)]\Bigg(\frac{1}{q^2-m_{h_1}^2}\Bigg) [\bar{u}(p_3)(y_{nh_1})_{22}u(p_1)]F(E_R)
\end{equation}

Where, $f_p$ and $f_n$ are the form factors of the proton and neutron, respectively. Squaring the matrix element and evaluating the traces using \texttt{FeynCalc} gives the following:
\begin{equation}\label{1.30}
|\mathcal{M}|^2 = \frac{[Zf_p + (A-Z)f_n]^2 (y_{nh_1})^2_{22}(t-m_N^2)(4m_T^2-t)}{(t-mh_1^2)^2}
\end{equation}

The differential cross-section obtained from this matches with that given in Ref.~\cite{Dutta:2020scq}:

\begin{widetext}
    \begin{equation}
    \label{1.24a}
    \frac{d\sigma}{dE_R} = [Zf_p + (A-Z)f_n]^2 \frac{(y_{nh_1})^2_{22}}{16\pi E_\nu^2} \frac{(m_N^2 + 2m_TE_R)(2m_T + E_R)}{(m_{h_1}^2 + 2m_TE_R)^2}F^2(E_R)
    \end{equation} 
\end{widetext}
Where, $m_N$, $Z$ and $A$ are the mass of the HNL, atomic number, and mass number of the target nuclei, respectively. $F(E_R)$ is the nuclear form factor (Helm's form factor as used previously) and $f_{p,n}$ are given by~\cite{Falk:1999mq}:

\begin{equation}\label{1.25}
\frac{f_{p,n}}{m_N} = \sum_{q=u,d,s}f_{T_q}^{(p,n)}\frac{f_q}{m_q} + \frac{2}{27}\Bigg(1-\sum_{q=u,d,s}f_{T_q}^{(p,n)}\Bigg)\sum_{q=c,b,t}\frac{f_q}{m_q}
\end{equation}
where, $f_{s,c,b,t}=0$, $f_{(u,d)}=(y_{(u,d)h_1})_{11}$, $f_{T_u}^{(p)} = (20.8 \pm 1.5) \times 10^{-3}$, $f_{T_d}^{(p)} = (41.1 \pm 2.8) \times 10^{-3}$, $f_{T_u}^{(n)} = (18.9 \pm 1.4) \times 10^{-3}$, and $f_{T_d}^{(n)} = (45.1 \pm 2.7) \times 10^{-3}$~\cite{Junnarkar:2013ac, Hoferichter:2015dsa, Crivellin:2013ipa, Alarcon:2012nr, Alarcon:2011zs}. Upon substituting these values, we obtain $f_p=(9.01935\times 10^{-8})m_T$ and $f_n=(9.20369\times 10^{-8})m_T$, where $m_T$ is in MeV and $(y_{uh_1})_{11}=(y_{dh_1})_{11}=5.0\times 10^{-3}$ are the benchmark values. 

The matrix element for calculating the decay of the HNL into $\nu_\mu$ and $h_1$ is as follows:

\begin{equation}
|\mathcal{M}| = [\bar{u}(p_2,0)(y_{nh_1})u(p_1,m_N)]
\end{equation}

Using two-body kinematics we obtain the decay width of the HNL into $\nu_\mu$ and $h_1$ in the HNL's rest frame as:

\begin{equation}\label{1.26a}
\Gamma_{N\rightarrow \nu_\mu h_1} = \frac{(y_{nh_1})^2_{22}m_N}{16\pi}\Bigg(1-\frac{m_{h_1}^2}{m_N^2}\Bigg)^3
\end{equation}

Similarly, the decay width of $h_1$ into $e^{+}$ and $e^{-}$ in $h_1$'s frame, is given by:
\begin{equation}\label{1.27}
\Gamma_{h_1 \rightarrow e^{+} e^{-}} = \frac{(y_{eh_1})^2_{11}m_{h_1}}{8\pi}\Bigg(1-\frac{4m_e^2}{m_{h_1}^2}\Bigg)^{3/2}
\end{equation}

\section{Signal Kinematics for the MiniBooNE-excess Solution}\label{app:MiniBooNEExcessKinematics}
\begin{figure}[!htbp]
    \centering
    \begin{subfigure}{1.0\textwidth} 
        \centering
        \begin{subfigure}{0.48\textwidth}
            \centering
            \includegraphics[scale=0.50]{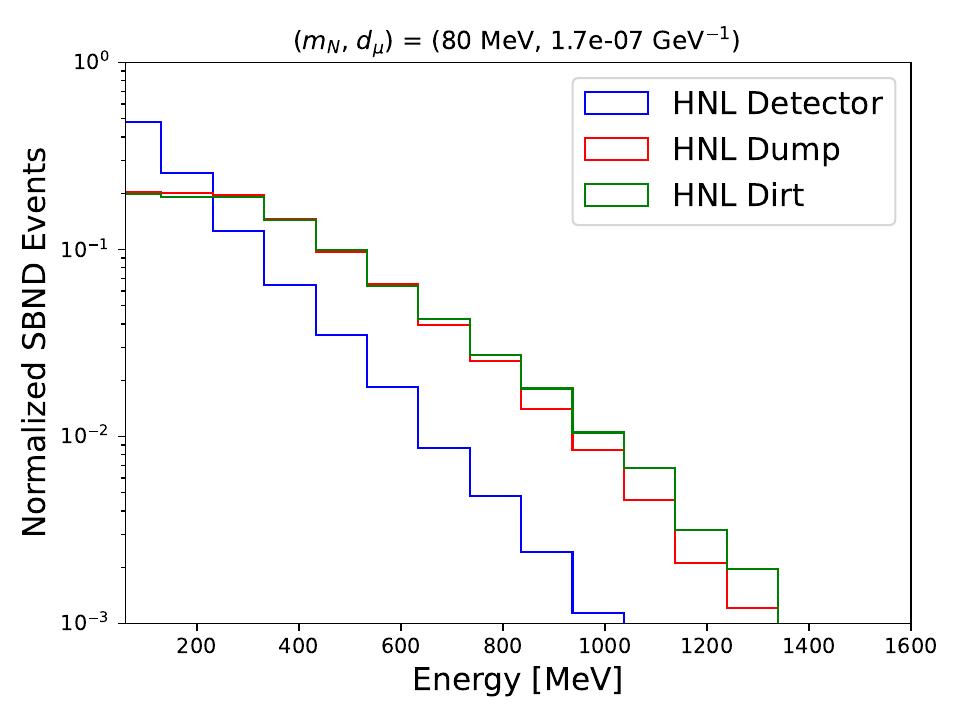}
        \end{subfigure}
        \begin{subfigure}{0.48\textwidth}
            \centering
            \includegraphics[scale=0.50]{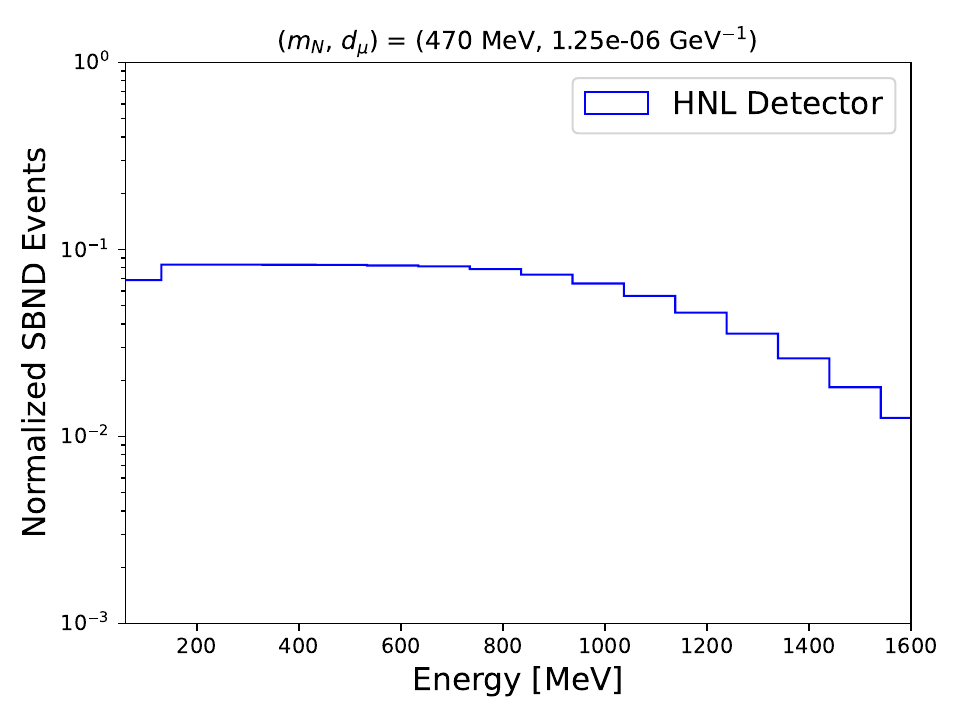}
        \end{subfigure}
    \end{subfigure}
    \begin{subfigure}{1.0\textwidth} 
        \centering
        \begin{subfigure}{0.48\textwidth}
            \centering
            \includegraphics[scale=0.50]{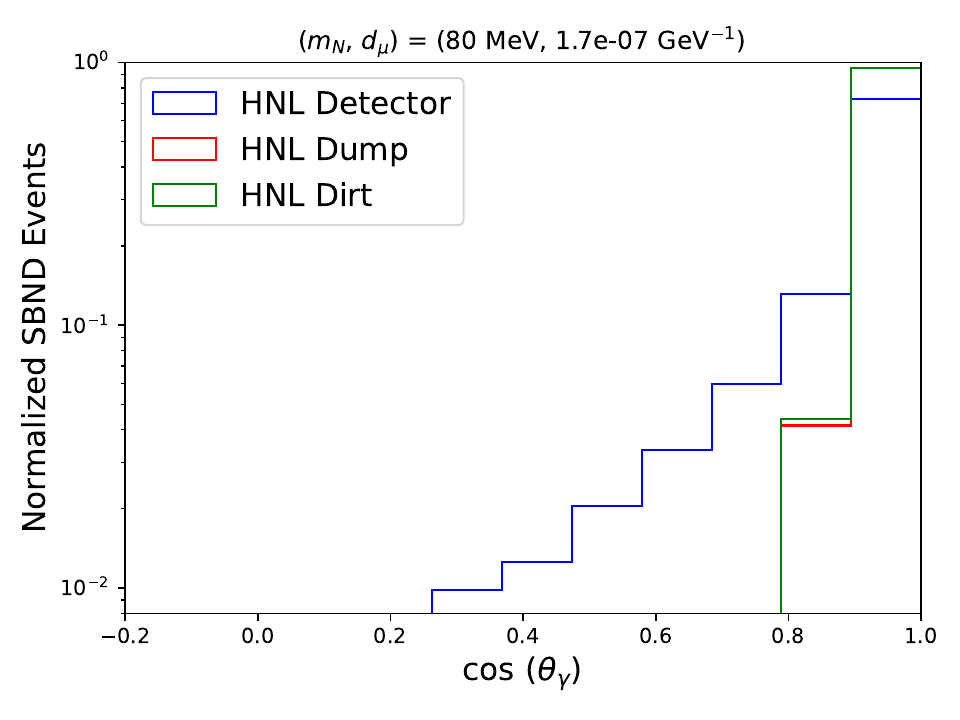}
        \end{subfigure}
        \begin{subfigure}{0.48\textwidth}
            \centering
            \includegraphics[scale=0.50]{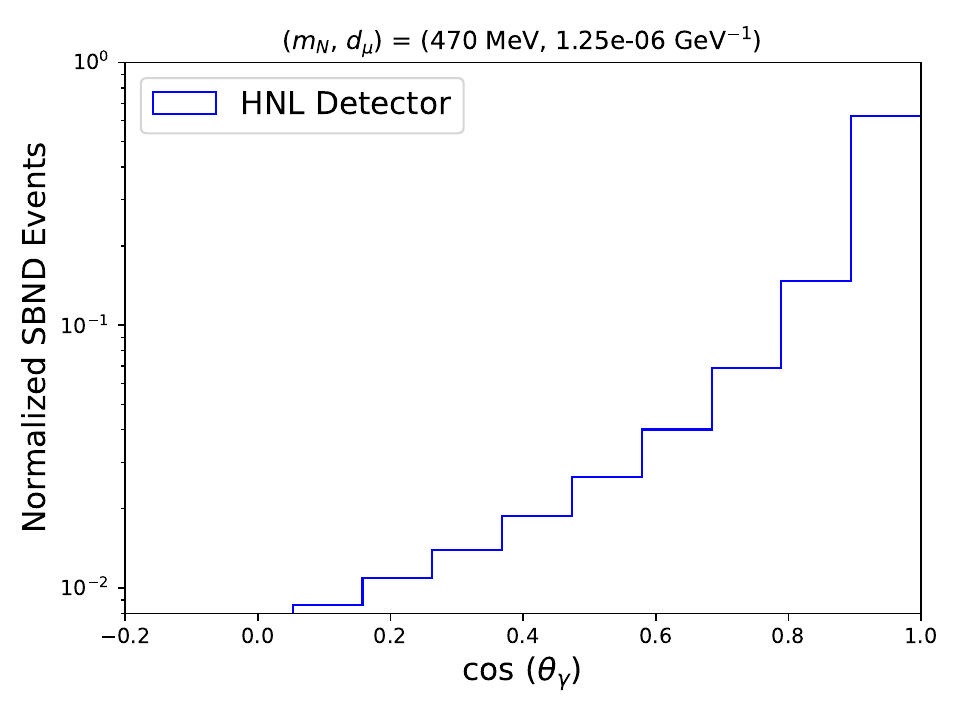}
        \end{subfigure}
    \end{subfigure}
    \captionsetup{justification=Justified,singlelinecheck=false}     
    \caption{Energy and Angular Spectra of the outgoing photons produced from decaying of the dpHNLs in the SBND detector. The mass and coupling are taken as the benchmark points for the MiniBooNE explanation in \cite{Kamp:2022bpt, Vergani:2021tgc}}
    \label{MiniBooNE_excess}
\end{figure}

In Figure \ref{MiniBooNE_excess}, we examine the energy and angular spectra that SBND will observe at the benchmark points discussed in Refs.~\cite{Kamp:2022bpt, Vergani:2021tgc}. These benchmark points, $(m_N, d_\mu)$: $(0.47~\text{GeV}, 1.25\times10^{-6}~\text{GeV}^{-1})$ and $(0.08~\text{GeV}, 1.7 \times 10^{-7}~\text{GeV}^{-1})$, are the preferred parameters that explain both the energy and angular MB excess. Since SBND is not sensitive to heavy dpHNLs ($m_N= 470$~MeV) produced at the dump and dirt, the plots to the right lack these lines.

Conversely, the left plots for lighter dpHNLs ($m_N= 80$~MeV) exhibit similar energy and angular characteristics as explained in sub-sections~\ref{sub:energy} and \ref{sub:angular}, with dump and dirt signals being more forward and energetic. In order to determine the exact number of events, we must perform a detailed background analysis, which is beyond the scope of this work. This demonstrates that SBND can not only probe the parameter space explaining the MiniBooNE anomaly but also observe additional features. 

\section{Additional Sensitivity-plot Comparisons for SBN Detectors}\label{app:MoreSensitivity}

\begin{figure}[!htbp]
\centering
\includegraphics[scale=0.45]{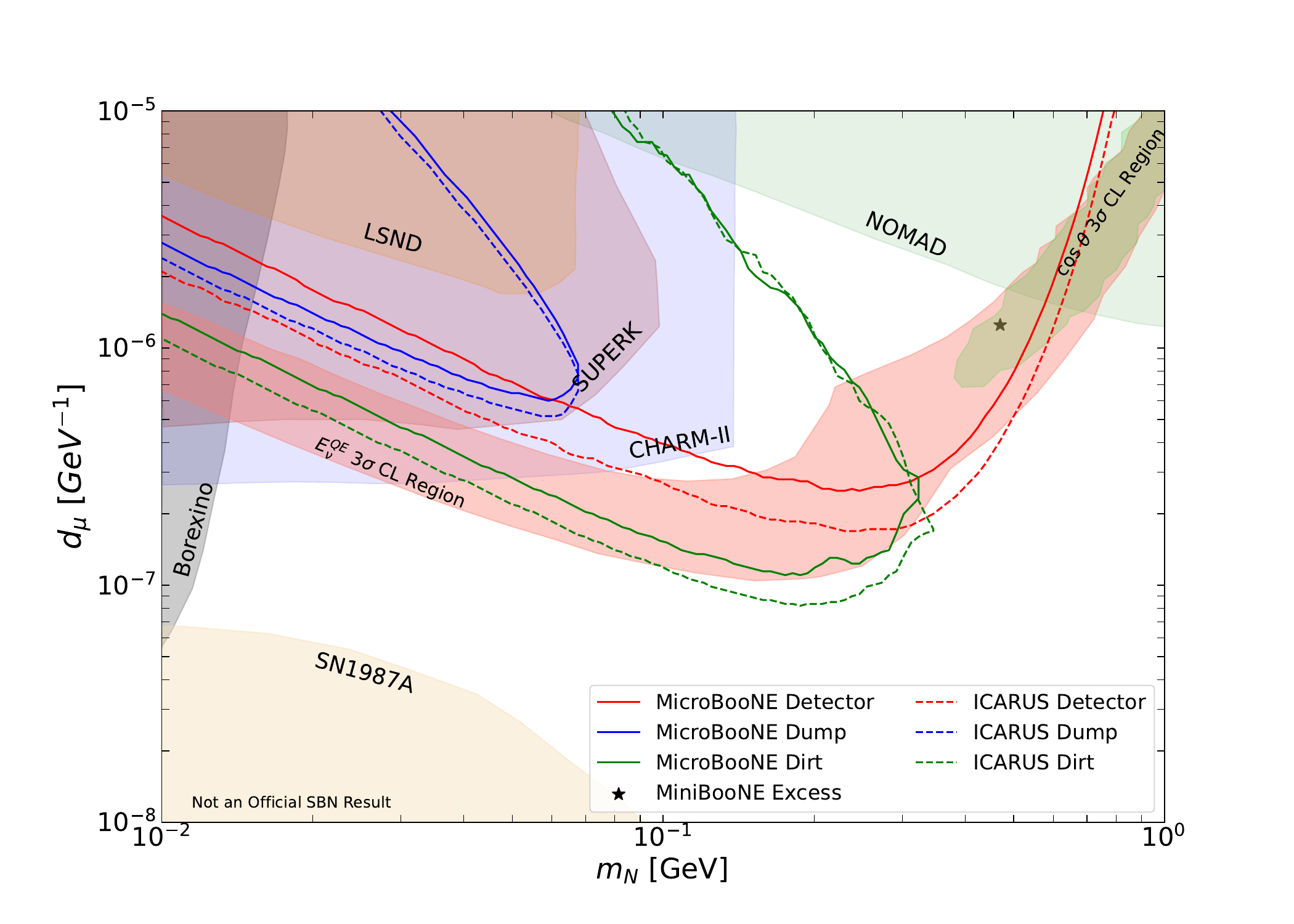}
\captionsetup{justification=Justified,singlelinecheck=false}
\caption{Sensitivity for individual production points of dpHNL in MicroBooNE and ICARUS. All the plots show 90\% confidence interval lines.} 
\label{ICARUS_MicroBooNE}
\end{figure}

In this Appendix, we show the sensitivity plots for MicroBooNE and ICARUS in Fig.~\ref{ICARUS_MicroBooNE} using the specifications and methodology mentioned in the main text. 

\begin{figure}[!htbp]
    \centering
    \includegraphics[scale=0.45]{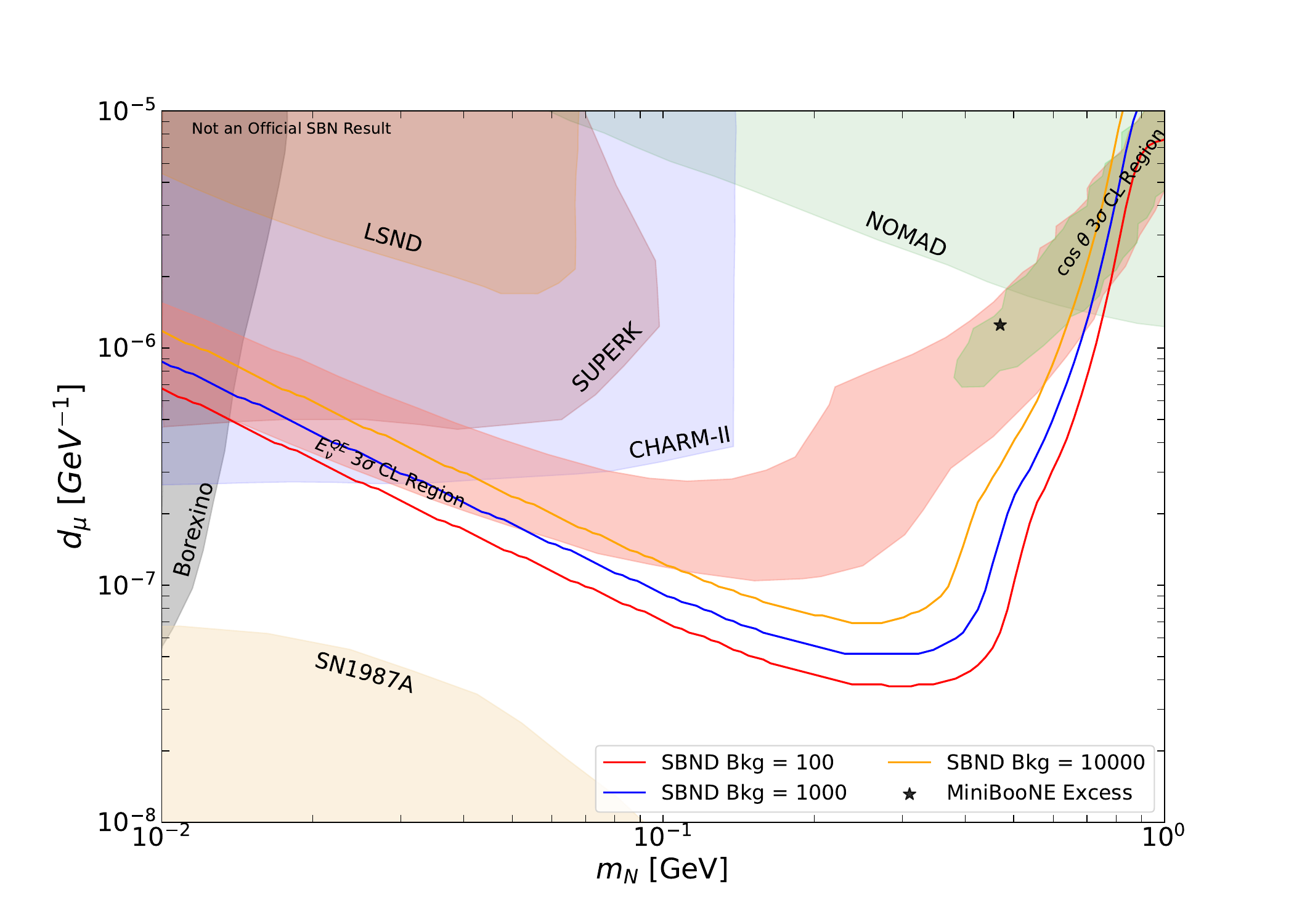}
    \caption{The 90\% Confidence Interval total contribution lines for SBND are shown for background levels of 100 (red), 1,000 (blue), and 10,000 (orange) events.} 
    \label{all_bkg}
\end{figure}

\section{Energy, Angular and Timing Spectra for ICARUS}\label{app:MoreComparisons}

\begin{figure}[!htbp]
    \centering
        \includegraphics[width=\textwidth]{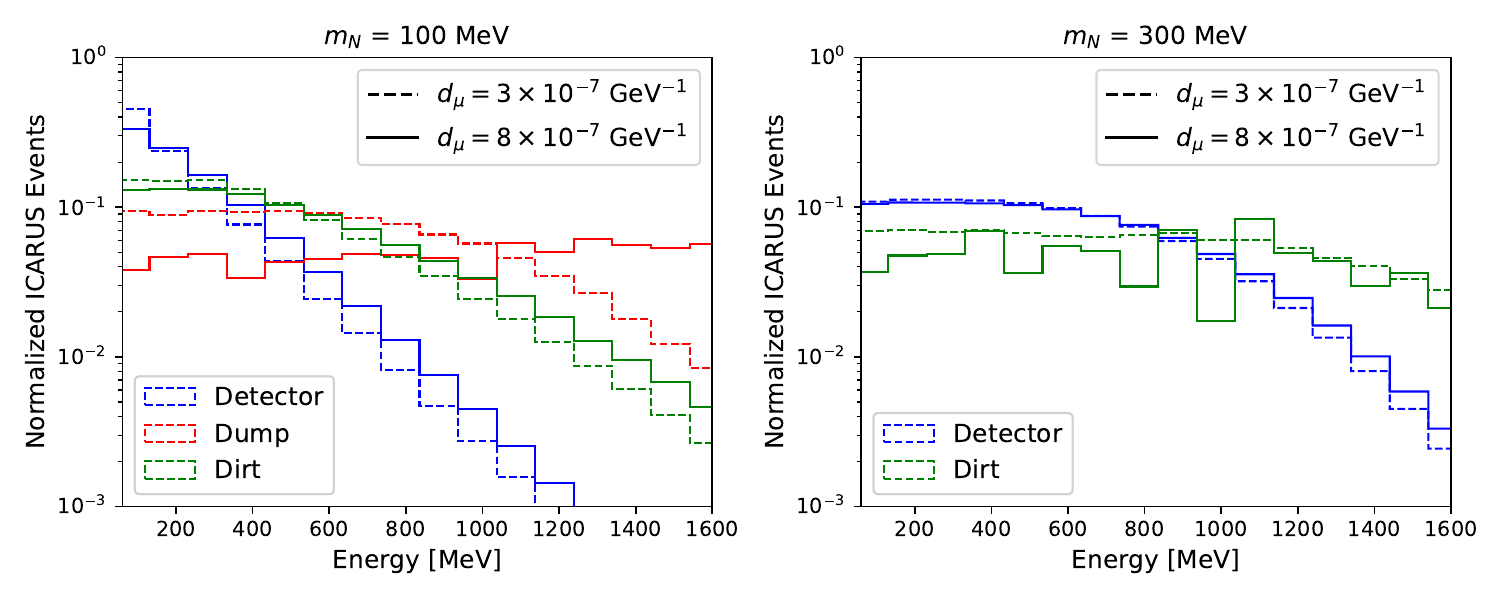}
    \captionsetup{justification=Justified, singlelinecheck=false}
    \caption{Energy spectra of outgoing photons produced by the decay of dpHNLs in the ICARUS detector. Each colored line corresponds to a different dpHNL production location. The x-axis represents the energy of the outgoing photon in the lab frame, while the y-axis shows the normalized number of ICARUS events per energy bin. Dashed and solid lines indicate values of $d_\mu=3 \times 10^{-7}$~GeV$^{-1}$ and $d_\mu=8 \times 10^{-7}$~GeV$^{-1}$, respectively. The left and right plots correspond to dpHNL masses of $m_N=100$~MeV and $300$~MeV, respectively.}
    \label{fig:EnergySpectrumICARUS}
\end{figure}

\begin{figure}[!htbp]
    \centering
    \includegraphics[width=\textwidth]{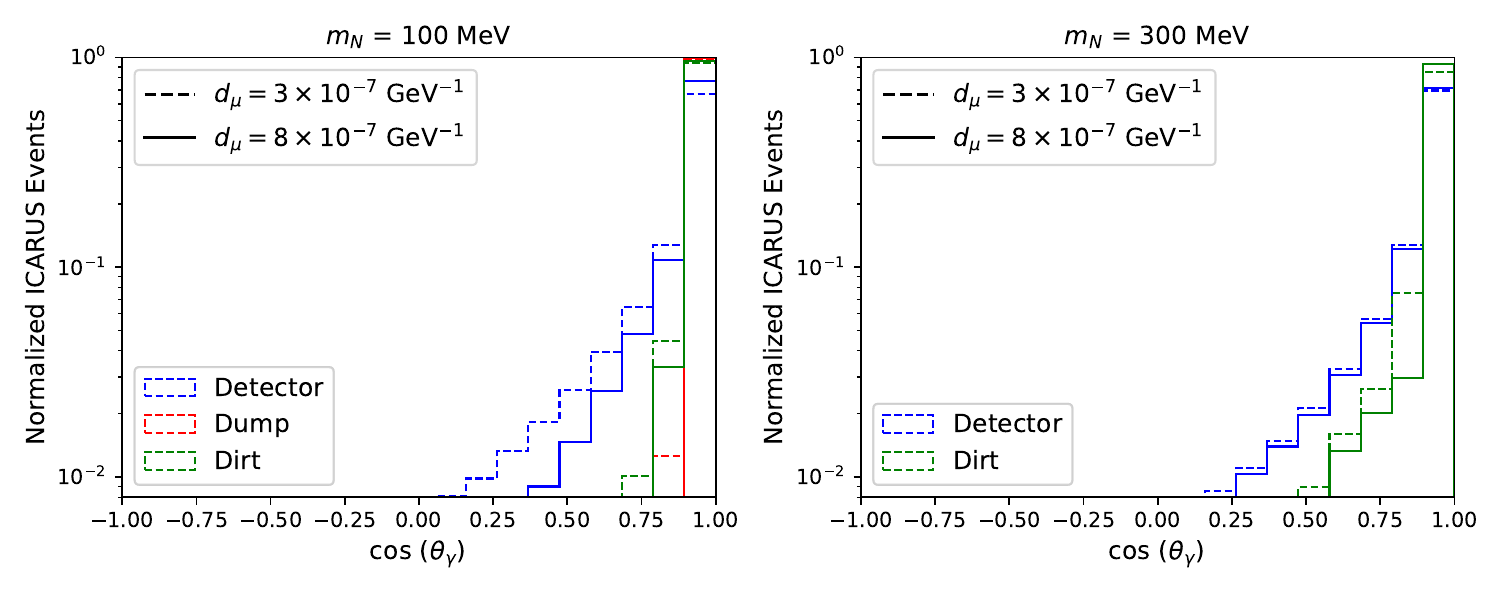}
    \captionsetup{justification=Justified, singlelinecheck=false}
    \caption{Angular spectra of outgoing photons produced by the decay of dpHNLs in the ICARUS detector. Each colored line corresponds to a different dpHNL production location. The angle $\theta_\gamma$ is defined by the outgoing photon direction relative to the beam axis. Dashed and solid lines indicate values of $d_\mu=3 \times 10^{-7}$~GeV$^{-1}$ and $d_\mu=8 \times 10^{-7}$~GeV$^{-1}$, respectively. The left and right plots correspond to dpHNL masses of $m_N=100$~MeV and $300$~MeV, respectively.}
    \label{fig:AngularSpectrumICARUS}
\end{figure}

\begin{figure}[!htbp]
    \centering
    \begin{subfigure}{0.48\textwidth} 
        \centering
        \includegraphics[scale=0.57]{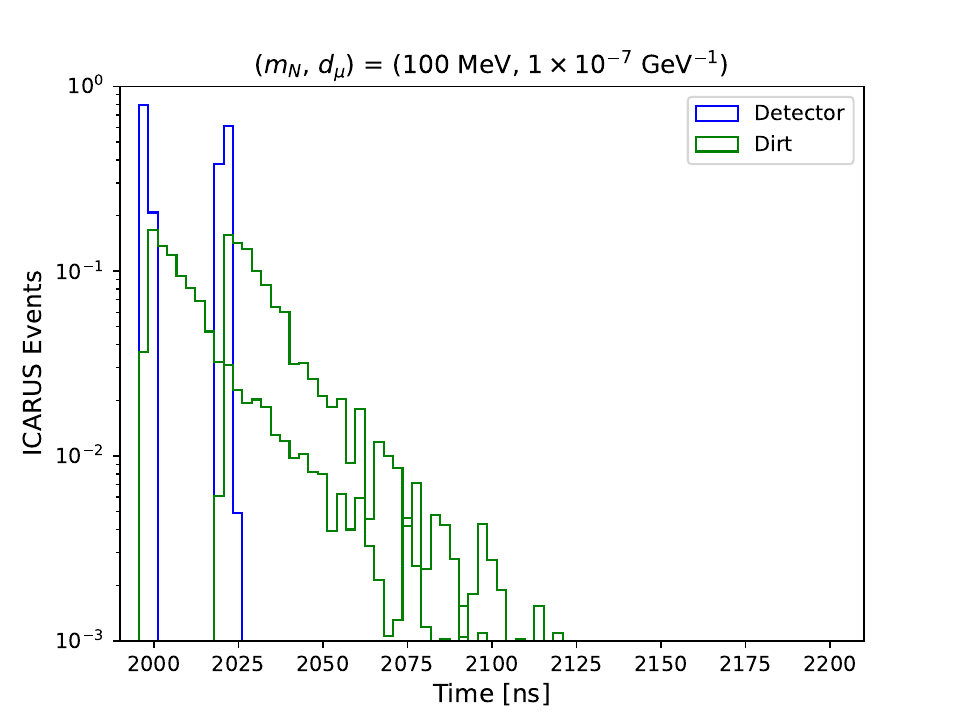}
    \end{subfigure}
    \begin{subfigure}{0.48\textwidth} 
        \centering
        \includegraphics[scale=0.57]{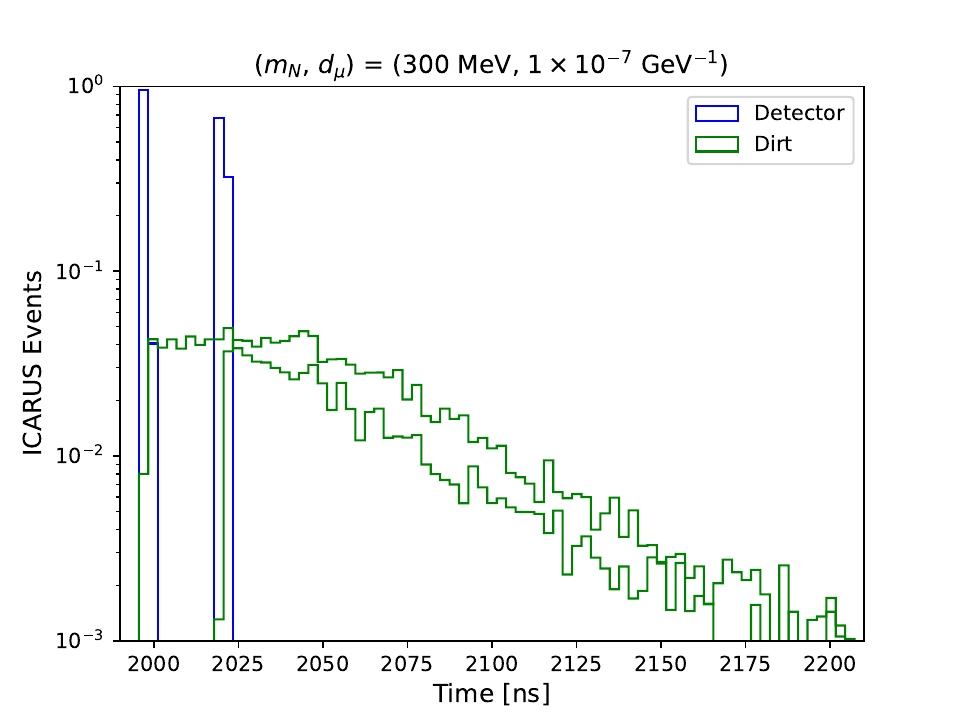}
    \end{subfigure}
    \captionsetup{justification=Justified, singlelinecheck=false}   
    \caption{Timing spectra of dpHNLs reaching the front face of the ICARUS detector for the dump and dirt lines, and $\nu_\mu$ for the detector lines. Here, $t=0$ represents the moment when the charged mesons cross the end of the magnetic horn. Only dpHNLs that produce visible signals in the detector are considered. The time on the x-axis represents the total time from $t=0$ until the dpHNLs/$\nu_\mu$ reach the ICARUS detector. }
    \label{fig:TimingSpectrumICARUS}
\end{figure}

In this Appendix, we present the energy, angular, and timing spectra for the ICARUS detector in Figs.~\ref{fig:EnergySpectrumICARUS},~\ref{fig:AngularSpectrumICARUS}, and~\ref{fig:TimingSpectrumICARUS}, respectively, to facilitate comparison with the corresponding spectra for SBND. Since ICARUS is located further downstream along the beamline, the spectra exhibit sharper peaks at higher energies, more forward angular distributions, and longer arrival times. Other features remain consistent with those described for SBND in \cref{sec:Distributions}. The absence of dump lines (red) for $m_N=300~$MeV is due to negligible sensitivity at ICARUS from dpHNLs produced in the dump contribution.

\bibliography{HNL}

\end{document}